\documentclass[journal abbreviation, manuscript]{copernicus}
\nolinenumbers

\usepackage{graphicx}
\usepackage[space]{grffile}
\usepackage{latexsym}
\usepackage{textcomp}
\usepackage{longtable}
\usepackage{tabulary}
\usepackage{booktabs,array,multirow}
\usepackage{amsfonts,amsmath,amssymb}
\usepackage{natbib}
\usepackage{url}
\usepackage{hyperref}
\usepackage{array}
\usepackage{adjustbox}
\usepackage{subcaption}
\usepackage{makecell}
\usepackage{xcolor} % color option
\hypersetup{
 unicode=false,          % non-Latin characters in Acrobat bookmarks
 pdftoolbar=true,        % show Acrobat toolbar?
 pdfmenubar=true,        % show Acrobat menu?
 pdffitwindow=false,     % window fit to page when opened
 pdfstartview={FitH},    % fits the width of the page to the window
 pdftitle={My title},    % title
 pdfauthor={Author},     % author
 pdfsubject={Subject},   % subject of the document
 pdfcreator={Creator},   % creator of the document
 pdfproducer={Producer}, % producer of the document
 pdfkeywords={keyword1, key2, key3}, % list of keywords
 pdfnewwindow=true,      % links in new PDF window
 colorlinks=true,       % false: boxed links; true: colored links
 linkcolor=purple,          % color of internal links (change box color with linkbordercolor)
 citecolor=purple,        % color of links to bibliography
 filecolor=purple,      % color of file links
 urlcolor=purple           % color of external links
}
\usepackage{etoolbox}
\usepackage{float} % for table[H]

\newif\iflatexml\latexmlfalse
\AtBeginDocument{\DeclareGraphicsExtensions{.pdf,.PDF,.eps,.EPS,.png,.PNG,.tif,.TIF,.jpg,.JPG,.jpeg,.JPEG}}

\newcommand{\merra}{MERRA-2}
\newcommand{\ibtracs}{IBTrACS}

\usepackage{xspace}
\usepackage{xcolor}
\newcommand{\mName}{TCG-Net\xspace}

\begin{document}

\title{Deep Learning Reconstruction of Tropical Cyclogenesis in the Western North Pacific from Climate Reanalysis Dataset}

% \Author[affil]{given_name}{surname}
\Author[1][trongld@vnu.edu.vn]{Duc-Trong}{Le} %1
\Author[1][]{Tran-Binh}{Dang} %1
\Author[1][]{Anh-Duc}{Hoang Gia} %1
\Author[1][]{Duc-Hai}{Nguyen} %1
\Author[1][]{Minh-Hoa}{Tien} %1
\Author[1][]{Xuan-Truong}{Ngo} %1
\Author[3][]{Quang-Trung}{Luu} %2
\Author[3][]{Quang-Lap}{Luu} %1
\Author[3][]{Tai-Hung}{Nguyen} %1
\Author[1][]{Thanh T.N.}{Nguyen} %1
% \Author[2][ckieu@indiana.edu]{Chanh}{Kieu} %1

% \Author[1]{}{} %3
% \Author[2]{}{} %4
% \Author[1]{}{} %5
% \Author[2]{}{} %6
% \Author[3]{}{} %7
\affil[1]{Faculty of Information Technology, VNU University of Engineering and Technology, Hanoi, Vietnam}
\affil[2]{Department of Earth and Atmospheric Sciences, Indiana University, Bloomington, IN 47405, USA}
\affil[3]{School of Electrical and Electronic Engineering, Hanoi University of Science and Technology, Vietnam}
% \affil[3]{}

% \Author[affil]{given_name}{surname}
\Author[2][ckieu@iu.edu]{Chanh}{Kieu} %1
% \Author[]{}{}
% \Author[]{}{}
%\affil[]{Department of Earth and Atmospheric Sciences, Indiana University, Bloomington, IN 47405, USA}
% \affil[]{ADDRESS}
%\equalcontrib{1,2}

%% The [] brackets identify the author with the corresponding affiliation. 1, 2, 3, etc. should be inserted.

%% If an author is deceased, please add \deceased[$Deceased date if applicable$]{$Author number$} (e.g. \deceased[13 November 2015]{2}) at the end of the affiliations. The author number depends on the placement of the author in the author list, e.g. the third author has number 3.

%% If authors contributed equally, please add \equalcontrib{$Author numbers$} (e.g. \equalcontrib{1,3}) at the end of the affiliations. The author number depends on the placement of the author in the author list, e.g. the third author has number 3.

\runningtitle{Reconstructing Tropical Cyclogenesis Climatology using Deep Learning}

\runningauthor{TEXT}
\received{}
\pubdiscuss{} %% only important for two-stage journals
\revised{}
\accepted{}
\published{}

%% These dates will be inserted by Copernicus Publications during the typesetting process.
\firstpage{1}
\maketitle

\begin{abstract}
This study presents a deep learning (DL) architecture based on residual convolutional neural networks (ResNet) to reconstruct the climatology of tropical cyclogenesis (TCG) in the Western North Pacific (WNP) basin from climate reanalysis datasets. Using different TCG data labeling strategies and data enrichment windows for the NASA Modern-Era Retrospective analysis for Research and Applications Version 2 (MERRA-2) dataset during the 1980–2020 period, we demonstrate that ResNet can reasonably reproduce the overall TCG climatology in the WNP, capturing both its seasonality and spatial distribution. Our sensitivity analyses and optimizations show that this TCG reconstruction depends on both the type of TCG climatology that one wishes to reconstruct and the strategies used to label TCG data. Of interest, analyses of different input features reveal that DL-based reconstruction of TCG climatology needs only a subset of channels rather than all available data, which is consistent with previous modeling and observational studies of TCG. These results not only enhance our understanding of the TCG process but also provide a promising pathway for predicting or downscaling TCG climatology based on large-scale environments from global model forecasts or climate output. Overall, our study demonstrates that DL can offer an effective approach for studying TC climatology beyond the traditional physical-based simulations and vortex-tracking algorithms used in current climate model analyses.  
\end{abstract}

% \copyrightstatement{TEXT} %% This section is optional and can be used for copyright transfers.
%%%%%%%%%%%%%%%%%%%%%%%%%%%%%%%%%%%%%%
\introduction  %% \introduction[modified heading if necessary]
\label{sec:Intro}
%%%%%%%%%%%%%%%%%%%%%%%%%%%%%%%%%%%%%%
The western North Pacific (WNP) basin has been well-documented to be the most active area of tropical cyclone (TC) activities \citep{Kossin_etal2016, Peduzzi_etal2012, Duc_etal2020}. With favorable conditions for TC formation (also known as tropical cyclogenesis, or TCG) such as warmer sea surface temperature (SST), active monsoon trough formation, or frequent convectively coupled equatorial waves, WNP produces about 25-30 TCs annually with a quarter of that affecting Vietnam coastal regions \citep{DefforgeMerlis2017, Duong_etal2021, Thanh_etal2020}. From the climate perspective, any change in TC main characteristics such as frequency, genesis locations, or intensity is often considered to be a manifestation of climate change. Thus, developing effective methods to construct TC climatology from different climate datasets is of significant importance for studying future projections of TC climatology from climate model outputs or outlooks from global forecasting systems \citep[e.g.,][]{Knutson_etal1998, Walsh_etal2007, Bengtsson_etal2007, Lee_etal2020, Cha_etal2020, Camargo_etal2023, Kieu_etal2023}. 

Traditionally, creating a TC climatology from gridded climate output involves using a vortex tracking algorithm to detect TC centers. This algorithm relies on a set of TC characteristics, such as absolute vorticity, surface maximum wind, minimum central pressure, warm core, or lifetime, which are checked at each model grid point \citep{ Walsh_etal2015, Zhao_etal2009, Strachan_etal2013, CamargoZebiak2002, Horn_etal2014, ZarzyckiUllrich2017, UllrichZarzycki2017, Vu_etal2021}. While being effective for high-resolution model outputs where TC characteristics are well-captured and thus suitable for weather forecast models or well-developed TCs, vortex tracking methods face challenges with coarse-resolution climate models (>0.5$^\circ$). For these coarse-resolution models or datasets, TC characteristics, especially during early formation, are often unclear \citep{ZarzyckiUllrich2017, Tran_etal2020}. Consequently, directly tracking a vortex from such outputs can lead to uncertainties in the timing and location of early TCG. This issue becomes more serious when studying climate change aspects like shifts in TCG location or timing across different climate datasets \citep[e.g.,][]{Phan_etal2015, Tran_etal2020}.    

The rapid advancement of machine learning (ML) techniques has opened new avenues for atmospheric research as well as operational forecasting. Given the vast amount of observational and model-generated data, weather and climate systems naturally provide a "big data" platform that are well-suited for training ML models not only for short-term weather prediction but also for capturing long-term climate variability \citep[e.g.,][]{Schultz_etal2021, pathak_etal2022, Bi_etal2021, Remi_etal2023, NguyenKieu2024}. In fact, several private and governmental organizations have recently developed deep learning architectures that outperform traditional physics-based models in weather forecasting, as demonstrated by \citet{pathak_etal2022} and \citet{Remi_etal2023}. Among recent applications of ML to TC research, most efforts have been limited to short-term contexts such as weather forecasting, satellite retrieval, or diagnostic studies. For examples, \citet{Miller_etal2017,Wimmers_etal2019} developed a deep learning (DL) model with satellite data to train a convolutional neural network, which can categorize TC intensity based on different cloud patterns and satellite channels. This line of approach has been further advanced to help improve TC forecasts by integrating the tracking information and/or other reanalysis data, with some modest performance for nowcasting and diagnoses \citep[e.g.,][]{Gao_etal2018, Kim_etal2019, Chen_etal2020, Giffard_etal2020}.

Focusing specifically on TCG from climate data, \citet[][hereinafter NK2024]{NguyenKieu2024} recently evaluated several DL architectures and identified promising potential for early warning of TCG events in the WNP basin. By treating climate reanalysis data as multi-channel input, they formulated TCG detection as a classification problem and demonstrated reasonable forecast skill, even at a coarse spatial resolution of $1^{\circ} \times 1^{\circ}$. While the model’s predictive skill declines with increasing forecast lead time, their approach highlights an important capability of detecting TCG directly from climate model outputs. In particular, certain DL architectures can be used to identify both the location and timing of TCG events within a given domain, which is valuable for broader applications in TCG climatology or future projections. It is important to emphasize here the distinction between predicting and detecting TCG. While long-lead TCG prediction suffers from rapidly decreasing skill due to the inherent limits of predictability in tropical dynamics, detecting TCG from coarse-resolution climate data (also commonly known as downscaling) proves to be much more skillful and feasible. This is because detection relies primarily on identifying favorable environmental conditions at the time of genesis, making it a problem of climate downscaling rather than one constrained by the chaotic dynamics of the atmosphere. 

Although existing ML applications for TC research show some promises, they have focused so far mostly on short-term prediction or nowcasting, rather than on TC climatology. Specifically, the use of ML for constructing TC climatology remains relatively preliminary,\citep[e.g.,][]{ScherandMessori2019, Wang_etal2022, ChenYuan2024}. Key challenges include how to leverage ML to diagnose climate model outputs, identify and correct model biases, or construct robust statistics of large-scale climate features. Thus, the development of ML-based techniques for downscaling of TC climatology or distributions of any other extreme events from reanalysis data is largely unexplored, yet represents an important open direction for future applications of DL for climate research. Given the rapid advancements in DL techniques, the main objective of this study is to introduce a framework for reconstructing TCG climatology from climate reanalysis datasets. Specifically, we will extract key TCG characteristics such as frequency or spatial distribution from the NASA Modern-Era Retrospective analysis for Research and Applications, Version 2 (MERRA-2), instead of forecasting TCG as in NK2024. Our DL-derived TCG climatology can serve as an independent validation and complement to those obtained from traditional vortex tracking methods. Furthermore, any TC climatology obtained from MERRA-2 data can also be used as a reference for examining the change of TC climatology in future projections, which justifies our DL approach herein.     

The rest of this work is organized as follows. In Section~\ref{sec:Methodology-Data}, details of data pre-processing and our CNN algorithms are presented. An approach to generate and label TCG binary dataset for each application will also be discussed. Section~\ref{sec:DL-models} presents the detailed design of our DL pipeline in this study, along with DL experimental designs. Section~\ref{sec:Results} provides results and related discussions. Finally, a summary and concluding remarks are given in Section~\ref{sec:Conclusion}.

%%%%%%%%%%%%%%%%%%%%%%%%%%%%%%%%%%%%%%
\section{Methodology and Data}
\label{sec:Methodology-Data} 
%%%%%%%%%%%%%%%%%%%%%%%%%%%%%%%%%%%%%%
%%%%%%%%%%%%%%%%%%%%%%%%%%%%%%%%%%%%%%
\subsection{Input data}
\label{sec:Dataset-Overview}
%%%%%%%%%%%%%%%%%%%%%%%%%%%%%%%%%%%%%%
To make our work directly applicable to research in TC climate downscaling or projection, the MERRA-2 dataset \citep{Gelaro_etal2017} was used in this study. This dataset is an atmospheric reanalysis based on the Goddard Earth Observing System Model (GEOS-5, Version 5) data assimilation system \citep{Gelaro_etal2017}. Unlike the original MERRA, \merra\, employed a newer version of GEOS-5 that assimilated newer microwave sounders and infrared radiance, as well as other data types. In particular, all data collections from \merra\, are provided on the same horizontal grid, which has $576 \times 361$ points in the longitudinal and latitudinal direction, respectively (a resolution of $0.625\times 0.5\degree$ longitude-by-latitude grid), and interpolated to 42 standard pressure vertical levels. While the output collections of \merra\, are on the regular $0.625 \times 0.5^{\circ}$ that are relatively coarse for TC inner-core region, our focus in this study is on the climatology of TCG, which requires mostly environmental conditions at the meso to synoptic-scale. As such, 0.5-degree resolution data is sufficient for our purposes as discussed in NK2024.

\begin{table}[t!]
    \centering
\begin{tabular}{|m{3cm}|>{\centering}p{3cm}|>{\centering}p{3cm}|>{\centering}p{3cm}|}
%\hline 
% & \fnl & \merra & \ibtracs \tabularnewline
%\hline 
%LAT MIN & -90 & -90 & -90\tabularnewline
%\hline 
%LAT MAX & 90 & 90 & 90\tabularnewline
%\hline 
%LAT STEP & 1 & 0.5 & 0.01\tabularnewline
%\hline 
%LON MIN & 0 & -180 & 0\tabularnewline
%\hline 
%LON MAX & 360 & 180 & 360\tabularnewline
%\hline 
%LON STEP & 1 & 0.625 & 0.01\tabularnewline
%\hline 
%TIME MIN & 01.01.1999 & 01.01.1980 & 01.01.1890\tabularnewline
%\hline 
%TIME MAX & 31.12.2022 & 31.12.2022 & 31.12.2022\tabularnewline
%\hline 
%TIME STEP & 06:00:00 & 03:00:00 & 03:00:00\tabularnewline
%\hline 
%File format & GRIB2 & NetCDF4  & Csv\tabularnewline
%\hline 
%File split & By forecasts & By day & {*}No splitting{*}\tabularnewline
%\hline 
%File size & 20MB-40MB & 2.2GB-2.3GB & 306MB\tabularnewline
%\hline
\hline 
Variable & \merra & \ibtracs \tabularnewline
\hline 
LAT MIN & -90 & -90\tabularnewline
\hline 
LAT MAX & 90 & 90\tabularnewline
\hline 
LAT STEP & 0.5 & 0.01\tabularnewline
\hline 
LON MIN & -180 & 0\tabularnewline
\hline 
LON MAX & 180 & 360\tabularnewline
\hline 
LON STEP & 0.625 & 0.01\tabularnewline
\hline 
TIME MIN & 01/01/1980 & 01/01/1890\tabularnewline
\hline 
TIME MAX & 12/31/2022 & 12/31/2022\tabularnewline
\hline 
TIME STEP & 3 hours & 6 or 3 hours \tabularnewline
\hline 
File format & NetCDF4  & Csv\tabularnewline
\hline 
File split & By day & {*}No splitting{*}\tabularnewline
\hline 
File size & 2.2--2.3 GB & 306 MB\tabularnewline
\hline
\end{tabular}
    \caption{A list of variables and their corresponding ranges in the \merra\, and \ibtracs\, datasets used for developing DL models.}
    \label{tab:raw-data-insights}
\end{table}

Although several reanalysis datasets are available, MERRA-2 was selected for this study primarily due to two reasons: ($i$) its spatial and temporal resolution is suitable for detecting  TCG, and ($ii$) its data format is convenient for ML model development. Unlike some TC metrics such as intensity or accumulated energy that require detailed inner-core structure, TCG is a process that is largely governed by environmental conditions. With a resolution of 0.5$^\circ$ MERRA-2 can therefore capture these environmental conditions effectively for DL purposes. Note that the MERRA-2 dataset provides gridded atmospheric data that includes 11 key meteorological variables at standard pressure levels, spanning the global domain from $90^{\circ}$ S to $90^{\circ}$ N latitude (at $0.5^{\circ}$ resolution) and from $180^{\circ}$ W to $180^{\circ}$ E longitude (at $0.625^{\circ}$ resolution). The dataset is available from January 1, 1980, to December 31, 2022, with data sampled every 3 hours. Each daily file contains 8 time slices and is stored in NETCDF4 format, with a file size of approximately 2.2–2.3 gigabytes. In total, the full dataset amounts to roughly 25 terabytes, offering a sufficiently large and detailed input for training DL models. One limitation of the MERRA-2 dataset as compared to other reanalysis datasets is that this dataset contains a single resolution of $0.5 \times 0.625^{\circ}$, while other reanalysis datasets such as ERA5 provide higher resolution up to $0.25^{\circ}$ at an hourly interval. Using such higher resolution datasets is certainly an advantage, as it can help optimize ML models. However, most current global climate projection outputs are given on $0.5^{\circ}$ to $1^{\circ}$ resolutions. Thus, using $0.5^{\circ}$ data can help better demonstrate the usefulness and facilitate the applications of our DL models in reconstructing TC climatology from global climate outputs as designed. Yet, ERA5 is considered to be among the best for climate reanalysis and DL model development, whether ERA5 is better than MERRA-2 in terms of TC climatology at the $0.5^{\circ}$ resolution has not been demonstrated. In this regard, our choice of MERRA-2 can be considered as a pre-learning step, which can be refined with ERA5 or any other climate datasets in the future if needed. For the purpose of implementing and evaluating our DL model, the MERRA-2 data is therefore sufficient. 

%One specific issue with the MERRA-2 dataset for TC applications is that it does not contain meteorological information at the surface level. For weather features like TCs, the surface data such as sea-level pressure may contain important information about TC intensity or structure. For the TCG application, we note however that a low sea level pressure contour generally is not well defined and so such the missing of sea level pressure field will have a small impact overall on our TCG climatology reconstruction. 

Along with the use of \merra dataset for training, we also employed the International Best Track Archive for Climate Stewardship (IBTrACS) \citep{knapp2010ibtracs} to label all TCG events and locations, which contains global TC records. In the WNP basin, \ibtracs\ contains data from January 1, 1890 through present day, sampled every $3$ hours and archived in a single CSV file with a total size of approximately 300 megabytes. The fact that this \ibtracs\ dataset is structured with the same synoptic times as the \merra\ dataset is of importance, because it allows us to pair these two datasets for supervised DL. The details of these datasets as well as their corresponding variables are summarized in Table~\ref{tab:raw-data-insights}.

%%%%%%%%%%%%%%%%%%%%%%%%%%%%%%%%%%%%%%
\subsection{Data pre-processing}
%%%%%%%%%%%%%%%%%%%%%%%%%%%%%%%%%%%%%%
As Table~\ref{tab:raw-data-insights}, \merra\ and \ibtracs\ datasets have their own format, structure, and parameters, it is necessary to first synchronize these datasets before any DL model development can be carried out. For this data pre-process step, we convert the longitude axis of \merra\ from $[-180;+180]$ to $[0;360]$ to match with the coordinate definitions in \ibtracs, which is needed so that the $\mathtt{latitude}$ and $\mathtt{longitude}$ values of \ibtracs\ can be located properly on the \merra\ coordinate grids for data extraction. 
In our study, a domain of a size $[35^\circ S-70^\circ N] \times [60^\circ-220^\circ]$ in the Pacific Ocean is then extracted from the global \merra\ dataset. For each \merra\ file, a timestamp is output at an interval of 6 hours to match with the best track data (note that the original \merra\ dataset consists of daily data files at an interval of 3 hours). All details of these data outputs after pre-processing \merra\ and \ibtracs\ are provided in Table \ref{tab:overview-dataset-prep}. The corresponding pre-process workflow is thoroughly described in our Zenodo repository \citep{le2025zenodo}.    

\begin{table}[t!]
\centering
\begin{tabular}{|m{2cm}|>{\centering}p{2.5cm}|>{\centering}p{2.5cm}|>{\centering}p{2.5cm}|>{\centering}p{2.5cm}|}
\hline 
% \cline{1-3} \cline{2-3}
& \merra & \ibtracs \tabularnewline
\hline 
LAT MIN & -35 & 0\tabularnewline
\hline 
LAT MAX & 70 & 60\tabularnewline
\hline 
LAT STEP & 0.5 & 0.01\tabularnewline
\hline 
LON MIN & 60 & 100\tabularnewline
\hline 
LON MAX & 220 & 180\tabularnewline
\hline 
LON STEP & 0.625 & 0.01\tabularnewline
\hline 
TIME MIN & 01.01.1980 & 01.01.1890\tabularnewline
\hline 
TIME MAX & 31.12.2022 & 31.12.2022\tabularnewline
\hline 
TIME STEP & 6 hours & 6 hours \tabularnewline
\hline 
File format & NetCDF4 & CSV \tabularnewline
\hline 
Sample size & By forecasts & 1024 records\tabularnewline
\hline 
File size & 97MB & 4.0KB\tabularnewline
\hline 
\end{tabular}
\caption{Pre-processed data format and structures obtained from the \merra\ and \ibtracs\ datasets for ML training data requirement.}
\label{tab:overview-dataset-prep}
\end{table}

%%%%%%%%%%%%%%%%%%%%%%%%%%%%%%%%%%%%%%
\subsection{Supervised dataset design}
\label{subsec:Dataset-Design}
%%%%%%%%%%%%%%%%%%%%%%%%%%%%%%%%%%%%%%
Given the pre-processed data described above, the next step is to generate a labeled dataset for supervised ML models. Specifically for the study of TCG, we require a binary dataset that indicates whether or not a TCG event occurred in the WNP basin needed for the reconstruction of TCG climatology. The process of creating such a binary dataset is critical, as it must include not only positive TCG cases but also a well-designed set of negative samples. Such a requirement of both well-designed positive and negative samples is essential for ML models to effectively learn the distinct features between TCG and non-TCG conditions. On one hand, a strong contrast between positive and negative samples increases the likelihood that ML models will identify key patterns, thereby improving performance. Nevertheless, the selection of positive and negative samples must also align with practical applications and purposes in climate research. In fact, the criteria for labeling positive/negative TCG events vary depending on the specific type of TCG climatology or forecast objective under consideration as will be shown in the Section \ref{sec:Results}.     

In this study, we follow NK2024 and define a positive TCG event as the first time a storm was recorded in the best track data. One could also define a TCG as the first moment that a tropical depression stage is recorded to make sure TCG characteristics are well-defined for DL training \citep{KieuNguyen2024}. However, our choice of the first time that a TC was recorded in the best track herein has the benefit of training a DL model that can detect TCG earlier, and so we will use this definition to label a positive TCG event. 

With this definition of \textit{positive TCG labels}, we can scan through all TC track records and take the first recorded location of each storm to create a target output for a TCG event. Given the typical scale of TCs, the positive TCG domain is chosen to be a square box of size $18\times18^{\circ}$ centered on the first recorded TCG location, which is equivalent to roughly $33\times32$ grid points with the \merra\, $0.5^{\circ}$ resolution. Finally, all relevant information related to a TCG event including its longitudes, latitudes, date, and time was stored in a csv database to facilitate our data sharing and input to the DL interface.
% For the negative-labeled TCG data, the issue turns out to be much more subtle as discussed in Kieu and Nguyen (2024). This is because one can have several different ways to define a negative TCG event, depending on the context and application. From the practical perspective, there are essentially three different ways to generate negative-labeled TCG data. The first apparent choice for the negative-labeled data is to choose a fixed domain in an area of interest that we wish to predict whether a TCG event will occur on any given day (hereafter referred to as the fixed domain, or FD approach). For this negative sampling strategy, we choose a fixed area, e.g., located at $18\text{N}$ latitude and $110\text{E}$ longitude, and check if there is any TCG occurs within this fixed area every 6 hours. If there is no TC forming in this sampling area at a given time, the sampling is labeled as a negative TCG event for that time. This FD strategy guarantees that any TCG event will be applied to the domain of interest for real-time forecast, yet it produces a largely--imbalanced dataset as the TCG negative labels are much more than the TCG positive labels (i.e., most of the days have no TCG). In general, this FD strategy will be most useful for a question of whether or not environmental conditions are right for a TCG in one specific area of interest.  
Depending on the context and applications, one can have several different ways to define a \textit{negative TCG event}, as discussed in \cite{KieuNguyen2024}. From the practical perspective, we propose in this study two following sampling strategies.

The first strategy for generating negative-labeled TCG samples, referred to as the Past Domain (PD) strategy, uses temporal context to distinguish between positive and negative labels. Specifically, for each positive TCG event occurring at time $t$, samples within a past window from $t - k$ to $t$ are labeled as positive, capturing the precursors leading up to the cyclone formation. In contrast, all earlier samples, i.e., those from $t - k - 1$ and further back in time, are labeled as negative. This approach aims to answer the question of why a TC forms at a specific time but not earlier, as discussed in \citet{KieuNguyen2024}. A key advantage of this strategy is that it preserves the geographical location between positive and negative samples while introducing temporal separation. However, depending on the chosen value of $k$, there may still be some overlap in favorable environmental conditions between the positive and negative samples. Moreover, as more historical data are labeled as negative, the dataset can become increasingly imbalanced.

The second approach for generating negative TCG data is referred to as the Dynamic Domain (DD) strategy. In this approach, for each positive TCG event at time $t$, we consider the surrounding spatial domains as negative samples. Specifically, one considers the eight adjacent regions to the positive TCG location, including the north-west, north, north-east, east, south-east, south, south-west, and west directions (see Figure~\ref{fig:ddm-relative-locations}). In this strategy, the eight surrounding domains are also shifted back in time, similar to the PD strategy. That is, negative samples are selected from the eight spatially adjacent regions but at earlier times, such as $t - n$ with $n \geq k + 1$, to ensure a temporal gap with the positive window. This type of domain selection for negative TCG data helps answer a question of why TCG occurs in one place but not in other places at the same day/time. One could randomly choose one of the eight negative domains to construct a more balanced binary dataset, as in \cite{KieuNguyen2024}, or choose all eight domains to increase the sample size. Unlike the PD approach, we note that this DM strategy will have a small chance of including a co-existing TC in the negative samples. This issue can be addressed by simply checking if there is any co-existing TC nearby within the negative TCG domain, and remove this domain or filter a TC out as discussed in \cite{Nguyen2023}. In this study, we use a simple approach of discarding all negative TCG domains that have a co-existing TC to avoid complications with changing environmental conditions due to vortex removal processes. This affects $<$ 1\% of all data points as we have in the WNP basin, as most TCG in this basin rarely overlap within a domain of an $18^\circ \times 18^\circ$ area.  

\begin{figure}[t!]
    \centering
\begin{tabular}{|>{\centering}m{1.5cm}|>{\centering}m{1.5cm}|>{\centering}m{1.5cm}|}
\hline 
\texttt{NW}
north-west  & \texttt{N}

north  & \texttt{NE}

north-east \tabularnewline
\hline 
\texttt{W}

west  & \texttt{P}

\textbf{positive TCG} & \texttt{E}

east \tabularnewline
\hline 
\texttt{SW}

south-west  & \texttt{S}

south  & \texttt{SE}

south-east \tabularnewline
\hline 
\end{tabular}
    \caption{Illustration of a TCG data labeling strategy based on the dynamical domain approach, for which a positive TCG label at one location is surrounded by 8 negative TCG labels for the sampling strategy.}
    \label{fig:ddm-relative-locations}
\end{figure}

As part of constructing the binary TCG dataset described above, it is important to note that TCs can form at any time of day, whereas the \ibtracs\, dataset provides records at fixed 6-hour intervals. Consequently, the actual genesis location of a TC may differ slightly from the position recorded in the best-track data. This spatial discrepancy typically depends on storm motion and the algorithms used for detecting vortex centers at operational centers. In this study, we assume that the TCG location does not vary significantly within a 6-hour window, which provides a reasonable basis for maintaining consistency in our binary TCG dataset design. This assumption also allows us to use the first recorded time in the \ibtracs\, dataset to define positively labeled TCG samples, which by convention are assigned a label of 1. All negative samples are labeled as 0.

Regarding undefined (NaN) values in the \merra\, dataset, we noticed a common issue of missing data at higher atmospheric levels. These missing values pose some challenges for both data interpolation and ML training. To address this, we exclude all data layers with substantial incompleteness and limit our analysis to layers below the 100-hPa level during the preprocessing stage. The potential impacts of these missing values on model performance will be further evaluated in the machine learning experiments presented in the following section.

%\begin{table}[htb]
%\adjustbox{max width=\columnwidth}{
%\begin{tabular}{|l|l|l|l|l|l|l|l|l|l|l|l|l|l|l|}
%\hline
%Variable                                                & PV    & H     & O3    & OMEGA & PHIS  & PS    & QI    & QL    & QV    & RH    & SLP   & T     & U     & V     \\ \hline
%\begin{tabular}[c]{@{}l@{}}Avg. NaN\\ ratio\end{tabular} & 0.015 & 0.010 & 0.015 & 0.015 & 0.000 & 0.000 & 0.015 & 0.015 & 0.015 & 0.015 & 0.000 & 0.015 & 0.015 & 0.015 \\ \hline
%\end{tabular}
%}
%\caption{\nan\ ratio in the \merra\ %dataset.}
%\label{tab:nan-ratio-merra2}
%\end{table}

\begin{figure}[t!]
  \centering
  % Panel (a)
  \begin{subfigure}[b]{0.99\textwidth}
    \includegraphics[width=0.9\columnwidth]{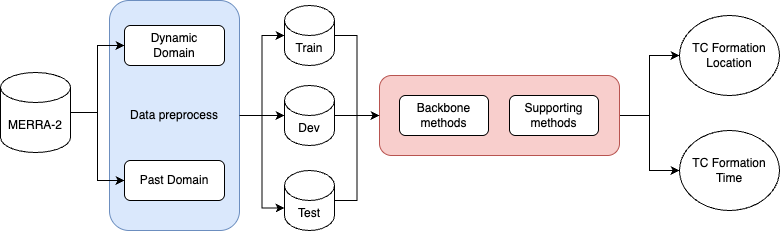}
    \caption{}
  \end{subfigure}
  \hfill
  % Panel (b)
  \begin{subfigure}[b]{0.99\textwidth}
    \includegraphics[width=0.9\linewidth]{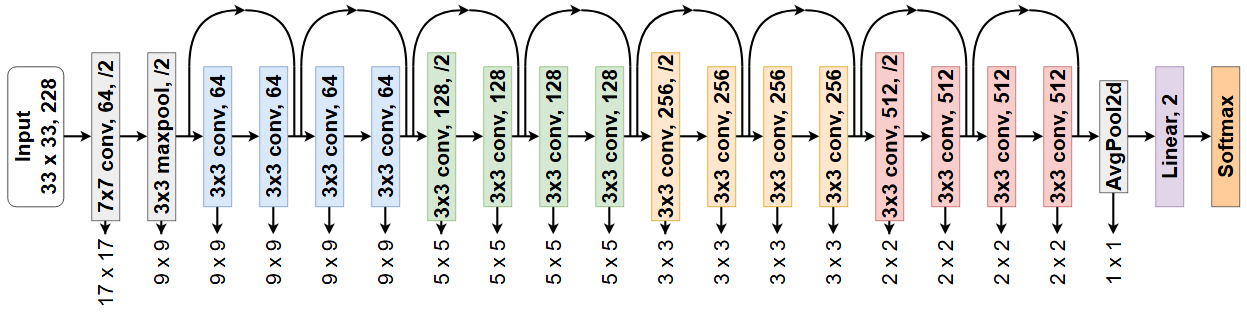}
    \caption{}
  \end{subfigure}
  \caption{a) The pipeline of the DL design for reconstructing TCG climatology from the MERRA-2 dataset, and (b) the core DL model based on the ResNet-18 architecture used for TCG reconstruction in this study.}
  \label{fig:full_pipeline}
\end{figure}

%%%%%%%%%%%%%%%%%%%%%%%%%%%%%%%%%%%%%%
\section{TCG-Net: Deep Learning Framework}
\label{sec:DL-models}
%%%%%%%%%%%%%%%%%%%%%%%%%%%%%%%%%%%%%%
\subsection{Model}\label{subsec:overview_method}
Recent advances in DL research have demonstrated substantial potential across a range of fields, including image recognition, natural language processing, autonomous driving, and weather prediction \citep{gao2018nowcasting, Kim2019, Giffard_etal2020, miller2017using, su2020applying, Bi_etal2021, Weyn_etal2021, NguyenKieu2024}. In the context of TCG applications, several DL approaches using satellite imagery data and TCG predictors have been proposed \citep{zhang2015discriminating, park2016detection, Matsuoka2018, Zhang2019, Kim2019}. To make further use of climate data in a broader context, NK2024 employed a convolutional neural network (CNN)-based DL framework for TCG application, which showed promising results despite limited training data. Their study underscores DL's image recognition capabilities, which enable the detection and analysis of relevant environmental patterns from climate meteorological fields for TCG prediction. Given such promising performance of the CNN-based models as well as the features tailored specifically for TCG reported in NK2024, we use a similar DL architecture based on the 18-layer Residual Net (ResNet-18) model in this study. ResNet-18 is a widely used CNN architecture for image recognition tasks. Its key advantage is the use of residual blocks, each consisting of two convolutional layers, a batch normalization layer, and a ReLU activation function, along with a `\textit{shortcut}' connection that directly adds the block's input to its output. This technique helps mitigate the `\textit{vanishing gradient}' problem, which commonly occurs in deep CNN architectures when the number of layers becomes excessively large.  
%
% This figure needs to be re-plotted to specifically reflect the DL design for the TCG problem herein, instead of a generic ResNet 18. Need to include also the data pre-process/post-process components to give readers an overview of how our entire pipeline looks like. 
%

In this study, we adapt the original ResNet-18 architecture for our problem of reconstructing the climatology of TCG. The modified network consists of eight residual blocks, preceded by an initial convolutional layer for input embedding, and followed by a fully connected layer with a softmax activation to predict the probability of storm occurrence, forming a total of 18 layers (Fig. \ref{fig:full_pipeline}b). This architecture differs slightly from that used in NK2024, as our study employs the MERRA-2 dataset, which lacks several environmental variables such as CAPE and tropopause-level features. Consequently, the ResNet-18 model required refinement and testing to achieve optimal performance, which differs somewhat from the architecture tailored for the NCEP Final reanalysis dataset in NK2024. The entire workflow for reading MERRA-2 data to final visualization is shown in Fig. \ref{fig:full_pipeline}a.
It is noted that our experiments with alternative DL architectures such as more convolutional layers, vision transformers, or pre-trained models all showed minimal improvement over the performance of the adapted ResNet-18 used in this study. While this observation is obtained from the few models that we have tried, it is possible that the TCG problem has limited predictability or MERRA-2 dataset may contain limited information for TCG at 0.5 degree resolution that more sophisticated models or architectures could not help learn further. This is a known problem in DL training, which explains why a simple CNN model could reach as good a performance as other models \citep{Brigato_etal2020}. Note also that complex models with more parameters generally have a very high "capacity" to learn. When the training data does not contain a variety of patterns, a high-capacity model can easily memorize the training data, including its noise and idiosyncrasies, instead of learning generalizable features. This leads to good performance on the training set but poor performance on unseen data (validation/test set), which is known as overfitting. Thus, the ResNet-18 model was adapted herein for our TCG reconstruction problem.     

%%%%%%%%%%%%%%%%%%%%%%%%%%%%%%%%%%%%%%
\subsection{Evaluation metrics}
\label{subsec:Experimental-Application}
%%%%%%%%%%%%%%%%%%%%%%%%%%%%%%%%%%%%%%
To monitor and verify our DL models, we employ several metrics from traditional classification problems as well as other meteorological metrics specific for TCG climatology such as seasonal or spatial distributions. For the DL model training, the performance of our DL model is based on Precision ($P$), Recall ($R$), and F1 score. These metrics are useful due to the inherent class imbalance, where TC occurrences (positive samples) are significantly fewer than non-TC cases (negative samples). By definition, Precision measures the proportion of correctly identified TC instances out of all instances predicted as TC:
\begin{equation}
    \text{Precision} = \frac{\text{TP}}{\text{TP} + \text{FP}},
\end{equation}
where TP (True Positives) represents correctly detected TCs, and FP (False Positives) denotes non-TC cases incorrectly classified as TC. A high precision ensures that the model minimizes false alarms, which is critical for reliable early warning systems.

In contrast, Recall quantifies the proportion of actual TC occurrences that the model successfully identifies:
\begin{equation}
    \text{Recall} = \frac{\text{TP}}{\text{TP} + \text{FN}},
\end{equation}
where FN (False Negatives) represents actual TCs that the model fails to detect. A high recall ensures that most TC events are captured, reducing the risk of missing critical storm formations.

To balance take into account both $P$ and $R$, we use the F1 score defined as:
\begin{equation}
    \text{F1-score} = 2 \times \frac{\text{Precision} \times \text{Recall}}{\text{Precision} + \text{Recall}}.
\end{equation}
A high F1 score indicates a good trade-off between false alarms and missed detections, making it a crucial metric for assessing the reliability of TC detection models. Since missed TCs can lead to severe consequences, while excessive false alarms may reduce trust in predictions, optimizing both Precision and Recall is essential in operational forecasting.

With the above verification metrics, we could train a DL model to maximize the model F1 score performance for any period, which is set by default to be 5 years of data from 2017-2023 for testing and 10\% of data from 1980-2016 for validation in this study (see Table \ref{tab:data}). Note that these category verification metrics are needed to evaluate the model's performance during training and testing. Whether the model can perform the required task depends further on its performance over other climate evaluations that are specific to the TCG problem such as seasonal distribution or spatial distribution, for which we will present in the next section.    

\begin{table}[t!]
    \centering
    \begin{tabular}{|c|c|c|}
         \hline
            Data subset & Period & Number of TCG events \\ \hline
         Training & 1980 - 2016 (Random 90\%) &1117   \\
         Validation & 1980 - 2016 (Random 10\%) & 124 \\
         Test & 2017 - 2023 & 188 \\ \hline
    \end{tabular}
    \caption{The default sampling ratio for training, validation, and test data based on the dataset MERRA-2 on experiments}
    \label{tab:data}
\end{table}

%%%%%%%%%%%%%%%%%%%%%%%%%%%%%%%%%%%%%%
\subsection{TCG data imbalance}
\label{subsec:TC-augmentation}
%%%%%%%%%%%%%%%%%%%%%%%%%%%%%%%%%%%%%%
Given the severe class imbalance of TCG data caused by the limited number of TCG events, i.e., $\sim$30 events per year in the WNP basin, it is essential to apply some additional techniques to mitigate this imbalance before training a DL model. One approach, as described in Section \ref{subsec:Dataset-Design}, involves temporal data enrichment using past windows, which is mostly suitable for the PD labeling strategy. This same past data window enrichment can also be extended to the DD labeling strategy by using the data from previous cycles to help increase the number of positive labels. Note however that the imbalance ratio for the DD labeling strategy is always 1:8, and so adding more past data would only help increase the overall number of TCG samples but not the imbalance ratio.   

%The first technique involves a data enrichment strategy for the PD labeling method. Specifically, for each TCG event recorded at time $T$ in the best track data, we label not only the data at $T$ but also data from $T$ to $T-k$ as positive TCG instances. Here, $k$ represents the number of preceding time intervals and referred to as the data enrichment window. This approach is motivated by the fact that, although a TC is officially recorded at time $T$, the environmental signals associated with its formation often emerge significantly earlier by 1-2 days. By including $k$ previous steps as positive TCG labels, we effectively amplify the number of TCG samples by $k$ times, thus capturing more relevant precursor signals.

In this study, we also introduce a complementary method to address class imbalance, which is the Random UnderSampling (RUS) approach. Specifically, the RUS method controls the ratio between negative and positive TCG samples in the training dataset. For example, a RUS value of 1:4 maintains four negative samples for every positive one. Since this undersampling strategy may still leave a slight imbalance, we further apply class weighting in the loss function to emphasize the minority class during training. This technique increases the loss penalty for misclassifying rare (positive) cases, encouraging the model to better capture them. Two types of class weights are used in this study: the balanced class weight (default) and the proportional class weight, defined as 1/5 RUS. In addition to the temporal data enrichment, the RUS-based sampling and class weighting strategies can improve the performance of our DL models for TCG reconstruction, surpassing the simple hyperparameter optimization based on the F1-score criterion.
% Together with the temporal data enrichment, the RUS-based sampling and class weighting strategies can enhance the performance of our DL models for TCG reconstruction, beyond the mere hyperparameter tuning based on the F1-score criterion. 

%%%%%%%%%%%%%%%%%%%%%%%%%%%%%%%%%%%%%%
\subsection{TCG detection}
\label{subsec:TC-detection-algo}
%%%%%%%%%%%%%%%%%%%%%%%%%%%%%%%%%%%%%%
With the output from our ResNet-18 model, we can finally produce a TC probability map for climate reconstruction. Given two different strategies for data labeling based on the DD and PD methods, detecting a TCG event and assigning it to a point in space and time require some specific details that we need to discuss further here. The negative labels are designed differently for different practical purposes. %That is, suppose a TCG event appears at time $t$ and location $P$ on the map, our question is: can this event be predicted based on past climate information at location $P$?  

For the PD strategy, it focuses on the temporal aspect of TCG prediction, which generates negative samples at the same location $P$ as positive samples. Thus, TCG detection for this approach is straightforward, as one can simply choose a location $P$ at time $t$, obtain data at that same location from previous times $t - i$ to time $t$, where $i$ represents previous time steps, and then generate TCG probability for that location $P$. With this strategy, one can obtain TCG probability for any area of interest, so long as the model is trained with the same data frames from earlier times $t-i$ up to time $t$. In this regard, PD can provide TCG information for any location at a given time $t$. 

With the DD strategy, note that one wants to extract the map information of TCG for the entire WNP basin at any given time $t$, which explicitly displays the spatial variation of TCG. So, our approach for reconstructing TCG for this DD strategy is to divide the entire WNP basin into a box of $5^\circ \times 5^\circ$, and apply the ResNet-18 model on each box separately. This way, we can detect TCG for all boxes simultaneously at time $t$. Assuming that a positive TCG event is found when TCG probability is larger than a certain threshold (e.g., 0.5), one can generate a map of all TCG locations at time $t$ as expected.   

Because of these different purposes for the PD and DD strategies, our evaluation and interpretation of the model performance for TCG climatology have to be based on separate metrics and criteria as presented in the section \ref{sec:Results}. 

%%%%%%%%%%%%%%%%%%%%%%%%%%%%%%%%%%%%%%
% \subsection{Code availability and user guide}
% \label{subsec:code_availability}
% %%%%%%%%%%%%%%%%%%%%%%%%%%%%%%%%%%%%%%
% The complete source code and pipeline of the DL framework used for reconstructing TCG climatology in this study are publicly available in \cite{le2025zenodo}. The repository includes all scripts for data preprocessing, model training, evaluation metrics (including sensitivity analysis), and visualization. Our code structure is designed in a way to support not only the reproducibility of results herein but also further experimentation with different climate datasets or TCG-related tasks such as operational forecasting.

% With this purpose, our code structure is set up to process the data loading and preparation separately (Fig. \ref{f1}), for which PyTorch is used to read meteorological datasets in the ".pt" and "netCDF" format from directories indicated in the CSV files that go with them (Fig. \ref{f2}). The output from this component is then fed into 2 different pipelines, depending on the PD or DD labeling strategy as shown in Figs. \ref{f3}-\ref{f4}. To make it self-contained, all model performance evaluation, post-processing, and visualization are carried out in each separate directory corresponding to the PD or DD strategy. A detailed user guide for running the full workflow can be found in our repository and so will not be repeated here. 
% % Structure of code repo and usage (All). See EVADE examples

%%%%%%%%%%%%%%%%%%%%%%%%%%%%%%%%%%%%%%
\section{Results}
\label{sec:Results}
%%%%%%%%%%%%%%%%%%%%%%%%%%%%%%%%%%%%%%
%%%%%%%%%%%%%%%%%%%%%%%%%%%%%%%%%%%%%%
\subsection{DL model benchmarking}
\label{subsec:Validation}
\begin{figure}[t!]
\begin{center}
\includegraphics[width=1.0\linewidth]{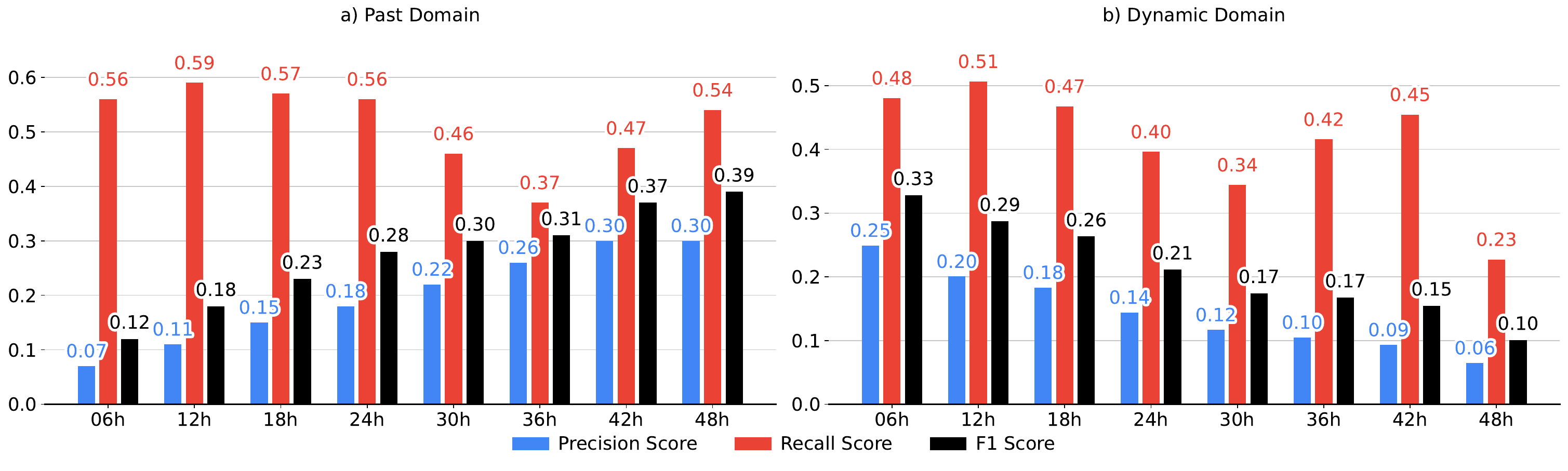}
\caption{Precision ($P$, blue columns), Recall ($R$, red columns), and F1 score (black columns) for the TCG prediction as obtained from the ResNet-18 model with the test set as a function of the temporal data enrichment window at an interval of 6 hours, using a) the PD sampling strategy, and b) the DD sampling strategy.}\label{fig:ResNet_optimization}
\end{center}
\end{figure}

%%%%%%%%%%%%%%%%%%%%%%%%%%%%%%%%%%%%%%
% To first have a general picture of how the ResNet-18 model is optimized for our TCG prediction problem, 
Fig. \ref{fig:ResNet_optimization} shows the precision, recall, and F1 score obtained from the PD and DD strategies. It is noted that for the PD strategy, we focus on a fixed domain over a part of the WNP basin whose TC activity has the most influence on Vietnam's coastal region ([0-30$^\circ$N]-[100-150$^\circ$W]). This domain choice is flexible and can be adjusted for each specific region in the operational forecast or climatology. Given such an arbitrary choice for the PD strategy, this section will report all results over the fixed domain.       

Clearly, the temporal feature enrichment plays a significant role in the performance of the ResNet-18 model. For the PD strategy, a longer period of feature enrichment generally gives a better precision score without much reduction in the recall performance (0.57-0.62), thus allowing for a higher F1 score for longer enrichment windows. Physically, this means that including more past information at one specific location generally helps improve the accuracy of the model performance in capturing TCG. However, the overall $P$ and $R$ scores are relatively low even when including all past information up to 48 hours prior to the formation of a TC. Thus, the F1 score is much less as compared to detecting TCG directly from satellite image data \citep[see, e.g.,][]{Chen_etal2020,Wimmers_etal2019}

As discussed in \cite{NguyenKieu2024}, such a low performance of DL models in detecting TCG from climate reanalysis datasets could reflect the fact that the TCG problem has limited predictability that including more information would not help improve its performance, consistent with the high false alarm rate in physical-based models. Another potential reason for such a low F1 score is the MERRA-2 data itself, which may not cover all possible environmental patterns for rare extreme events like TCG. This highlights the difficulty in reconstructing TCG climatology from climate reanalysis dataset.        

\begin{figure}[t!]
\begin{center}
\includegraphics[width=1\linewidth]{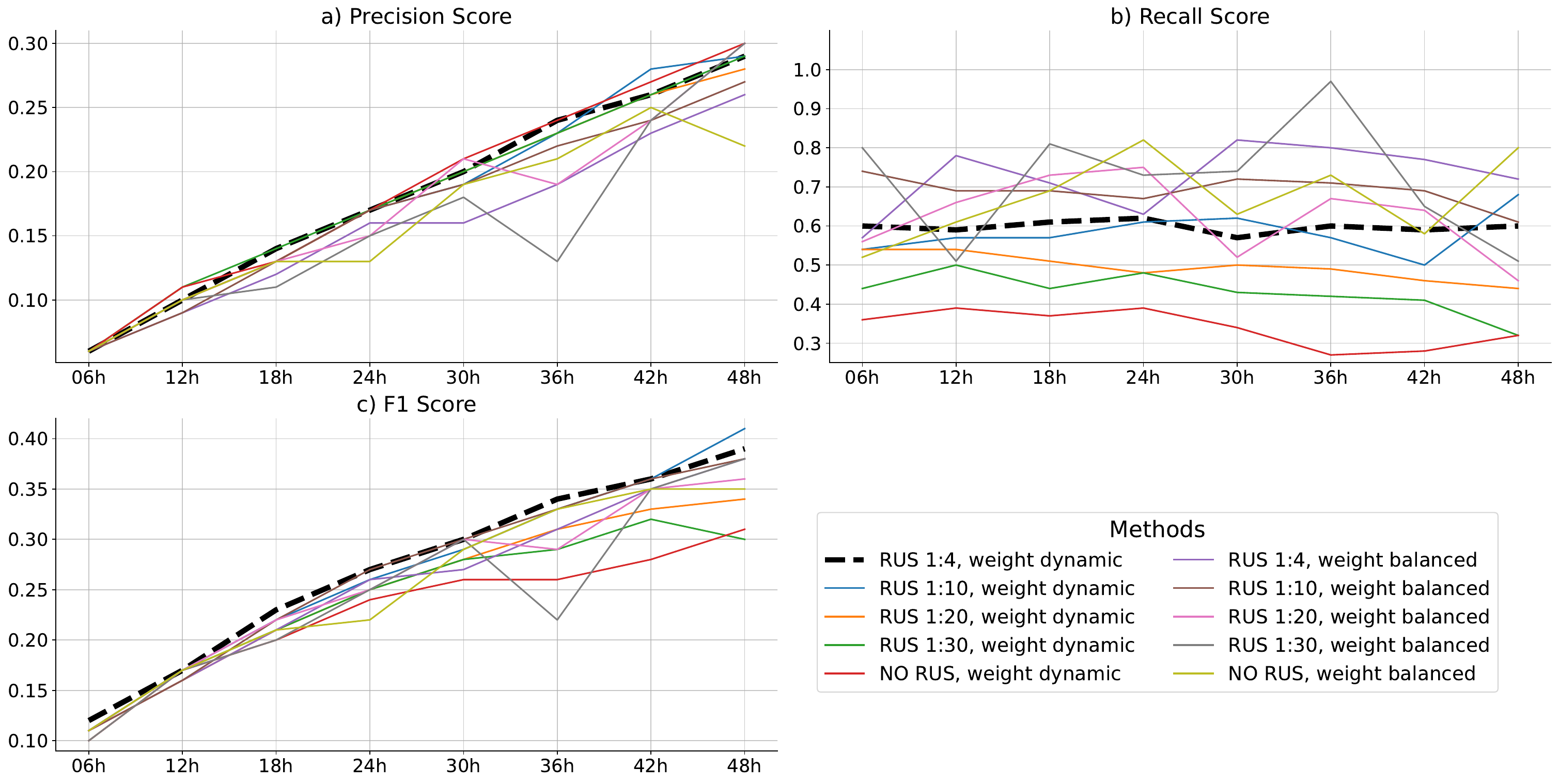}
\caption{(a) The $P$ score for the TCG prediction of the ResNet-18 model with the test set as a function of the temporal data enrichment window using a range of the RUS ratio and class weight (solid colors) for the PD sampling strategy, (b)-(c) similar to (a) but for the $R$ and F1 scores, respectively. Dashed black line denotes the values used for our best-tuned model. \textit{Weight balanced} assigns fixed importance to each class based on frequency whilst \textit{weight dynamics} adaptively adjusts sample or class importance.} \label{fig:RUS_sensitivity_PD}
\end{center}
\end{figure}

\begin{figure}[htb]
\begin{center}
\includegraphics[width=1\linewidth]{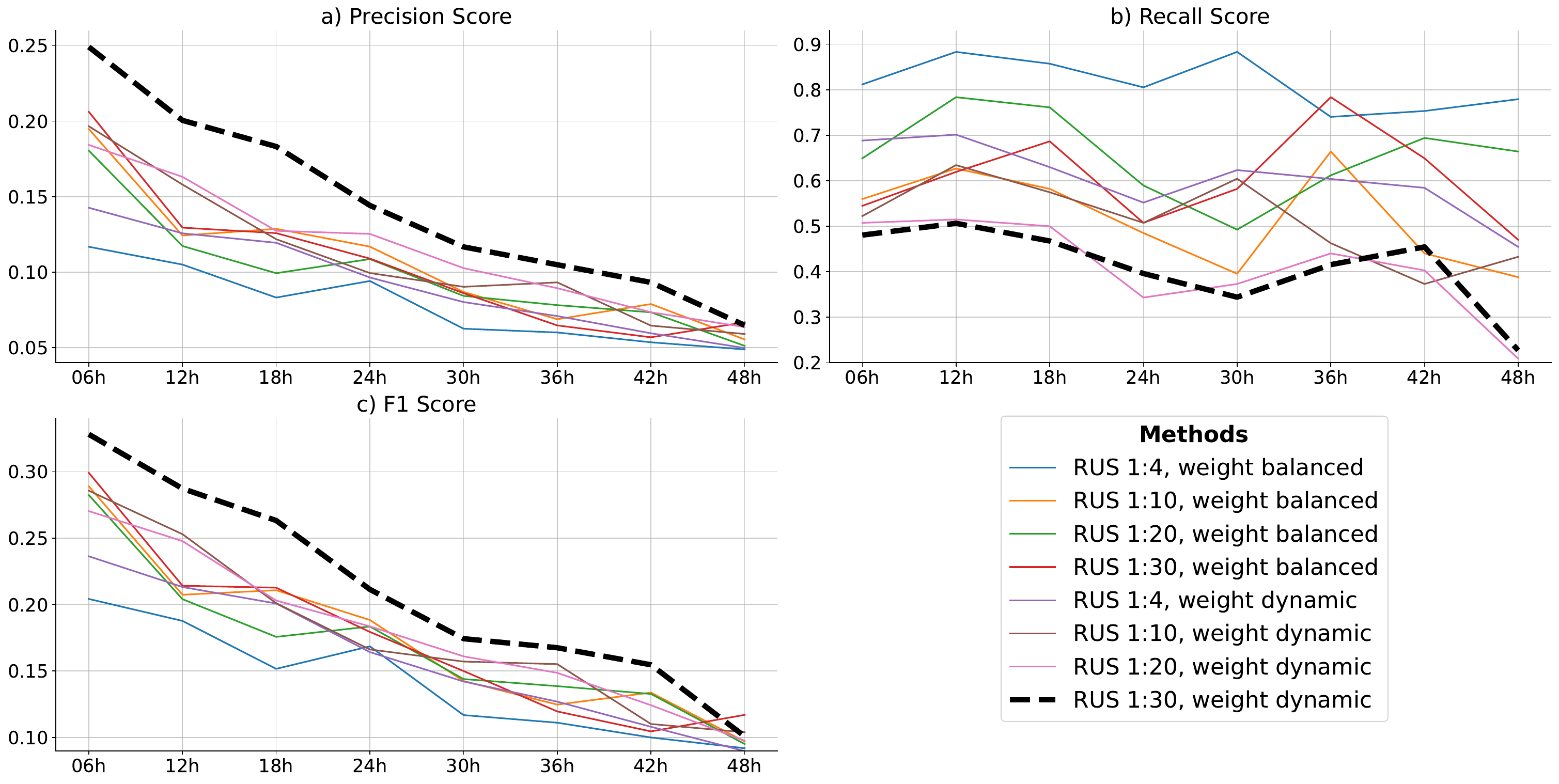}
\caption{Similar to Figure \ref{fig:RUS_sensitivity_PD} but for the DD sampling strategy.}
\label{fig:RUS_sensitivity_DD}
\end{center}
\end{figure}

For the DD strategy, the dependence of the model performance on the temporal enrichment is somewhat opposite. Specifically, the F1-score is optimal for very short enrichment during a 6 to 12-hour window, and it quickly decreases when the time window is longer. Practically, this means using past information to explain why a TC forms at one place but not at other places will not help if one includes more past information. This result is somewhat expected if one recalls that the main aim of the DD strategy is to search for where a TCG occurs at any given of time. So, the model is trained to detect TCG signals from the spatial information rather than from past information. Enriching the DD strategy by including more past information over the entire WNP basin therefore introduces more irrelevant environmental information from the past (i.e., negative labels) that causes some bias in the model towards negative samples. This explains why the $R$ score (blue columns in Fig. \ref{fig:ResNet_optimization}) decreases quickly with a longer enrichment window. As a result, the DD strategy exhibits a declining F1-score, indicating that the model's overall reliability in detecting TCG decreases when too much past information is used.

Given the fact that the TCG dataset is highly imbalanced due to the rarity of positive TCG labels, it is also important to examine how the ResNet-18 model could be optimized under various training data scenarios. For this purpose, Figure \ref{fig:RUS_sensitivity_PD} shows the ResNet-18 model performance with different RUS ratios as described in Section 3.3. For the sake of clarity, we show here the absolute RUS ratios with a fixed number of positive TCG labels (minority) while changing the number of negative labels (majority) according to each ratio displayed in Fig. \ref{fig:RUS_sensitivity_PD}. For each RUS ratio, a class weight value that adjusts the loss function is provided, which controls the importance towards the positive TCG labels during training.   

Figures \ref{fig:RUS_sensitivity_PD}-\ref{fig:RUS_sensitivity_DD} display a range of the $R$, $P$, and F1 scores for different RUS ratios and class weights, using both the PD and DD sampling strategies. One notices that a larger RUS ratio (i.e., more balance between positive and negative labels) tends to give higher $P$ and $R$ scores, thus resulting in a higher F1 score for the PD strategy. The optimal RUS ratio of 1:4 (one positive label corresponds to 4 negative labels) combined with a class weight of 0.5 provides the best performance for the PD strategy in terms of detecting TCG, which is chosen as a default value in Fig. \ref{fig:ResNet_optimization}. 

For the DD strategy, the model behavior is again somewhat different because of the constrain that one positive TCG location is surrounded by 8 negative labels by design. Therefore, the data imbalance at any given time is always fixed for the training. When applying the temporal enrichment with more data taken from the past for the test period, the number of negative TCG labels, however, increases rapidly because most of the days in the test period have no TCG. As a result, the imbalance ratio of the positive TCG over negative TCG labels becomes very small. For a reference, we show here only three RUS ratios of 1:10, 1:20, and 1:30 so one can compare their performance for the DD labeling strategy. 

Unlike the PD strategy, we notice that the DD strategy performs best when the RUS ratio is 1:30, with a weight class of 0.2. Too small or too large RUS ratios both degrade the model performance. Consistent with the control performance shown in Fig. \ref{fig:ResNet_optimization}, all RUS ratio and class weight experiments also show best performance when the data enrichment window is around 0-12 hours, beyond which the performance of the DD strategy starts decaying rapidly. Such behavior is likely because the inclusion of larger negative samples from surrounding environment at farther time back into the past cannot help the model distinguish the positive labels, leading to a much lower $R$ score. In constrast, too large RUS ratio means less over data for training. As a result, the overall F1 score is better for a RUS ratio of 1:30 and a shorter time window as seen in Fig. \ref{fig:ResNet_optimization}.      

In addition to the model optimizations based on the RUS ratio, feature enrichment time windows, and class weights as presented above, there are of course many other factors related to model architecture and hyperparameter settings that must also be considered to achieve optimal performance. Many of these aspects are excessively granular and so we cannot discuss every single one of them here. However, it is of interest to note that no single DL model is universally optimal across all weather features and spatial-temporal scales, particularly given the current limitations of available climate datasets and DL architectures.
% As such, the above results are specific for the data and the type of climate reconstruction.
% that we wish to present in this study.

In the context of the real-time TCG prediction task, we should note that optimizing a DL model for this task involves more than just the typical hyperparameter tuning. In fact, the optimization process is tied to the specific aim of predicting either the spatial distribution of TCG or the timing of TCG at one location, which corresponds to different data labeling strategies as presented above. Regardless of the strategies, our experiments with a range of model architectures from simple CNN to more complex DL models yielded similar performance in terms of the F1, precision, and recall scores (not shown). This observation supports our decision to fix the ResNet-18 architecture and focus on optimizing it with different model settings as presented herein, rather than exploring a broad array of alternative DL models for TCG climatology reconstruction purposes.

In the following section, we apply the best-performing model to reconstruct two important aspects of TCG climatology in the WNP basin, with the aim of ultimately extending this approach to climate downscaling in our future climate change studies.    

%%%%%%%%%%%%%%%%%%%%%%%%%%%%%%%%%%%%%%
\subsection{TCG reconstructed climatology}
\label{subsec:Validation}
%%%%%%%%%%%%%%%%%%%%%%%%%%%%%%%%%%%%%%

With the optimized ResNet-18 model for TCG climatology, our first attempt to reconstruct TCG climatology is the TCG seasonal distribution. For this purpose, the seasonal distribution is defined as the averaged ratio of the count of detected TCG events over the entire WNP basin in a given month to the total number of TCG events detected each year, which can be interpreted as the monthly TCG frequency. To avoid the arbitrary choice of the domain location in the PD strategy, we will generate this monthly TCG frequency for the DD strategy only.   

\begin{figure}[t!]
\begin{center}
\includegraphics[width=0.65\linewidth]{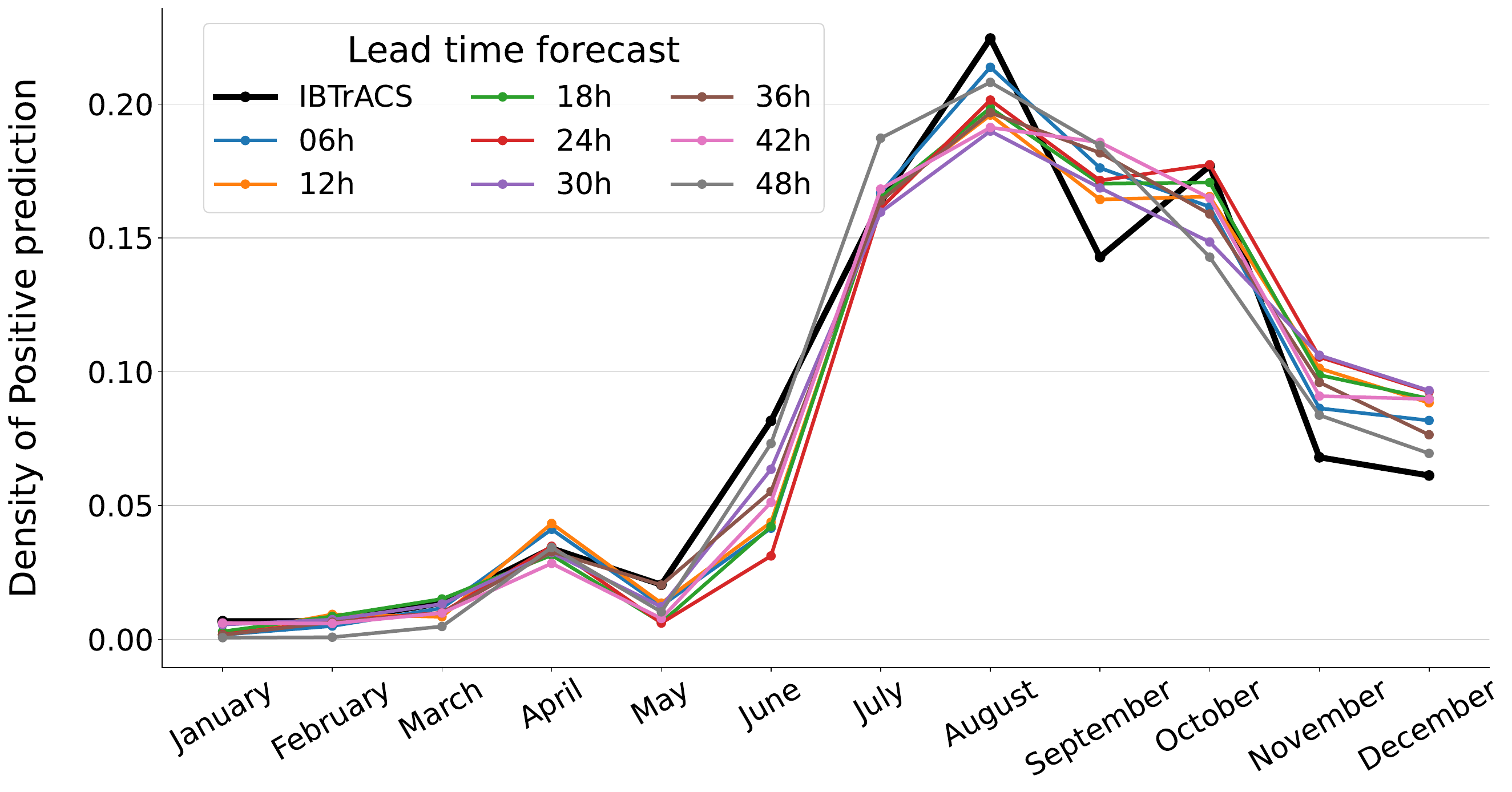}
\caption{Monthly distribution of the TCG frequency detected in the WNP basin from the test data (2017-2022), using the best-tuned ResNet-18 model for the DD strategy with 8 different time windows from 6 to 48 hours. The black solid curve denotes the TCG density obtained from the best track during the same time period.}
\label{fig:seasonalTCGdist}
\end{center}
\end{figure}

As seen in Figure \ref{fig:seasonalTCGdist}, the ResNet-18 model could reproduce well the overall seasonal distribution of TCG during the test period from 2017-2022, with a peak of TC activities in July-October, followed by an inactive period in January-April, similar to the observed TCG frequency. In particular, the ResNet-18 model could also reproduce the double peaks of TCG numbers in August and October consistent with the observed distribution, albeit the dip of TCG frequency in September is not as clear as that from the best track. During the peak period, note that the trainings with longer enrichment windows (i.e., 36-48 hours) tend to generate more positive TCG events than those with shorter windows (6-24 hour), suggesting that early signals of TC formation become more detectable as more past information is included. 

Towards the end of the peak season (during November-December), note that our DL model tends to produce more TCG than the observation, while it tends to underestimate the TCG frequency during May-June. These differences indicate a common fact in TC climate research that optimizing a DL model based on one specific metric such as F1 score, precision, or recall would lead to biases in other metrics \citep[See, e.g.,][]{Vu_etal2025}. Regardless of such differences due to different tunings, the consistency of the seasonal TCG reconstruction by the ResNet-18 model is seen across the time enrichment windows, thus providing confidence in the robustness of our model. This result also highlights the ResNet's broader ability to understand the seasonal changes in large-scale environments for TCG, which are needed for practical TCG prediction from global model outputs or climate change analyses. 

\begin{figure}[htb]
\begin{center}
\includegraphics[width=1.0\linewidth]{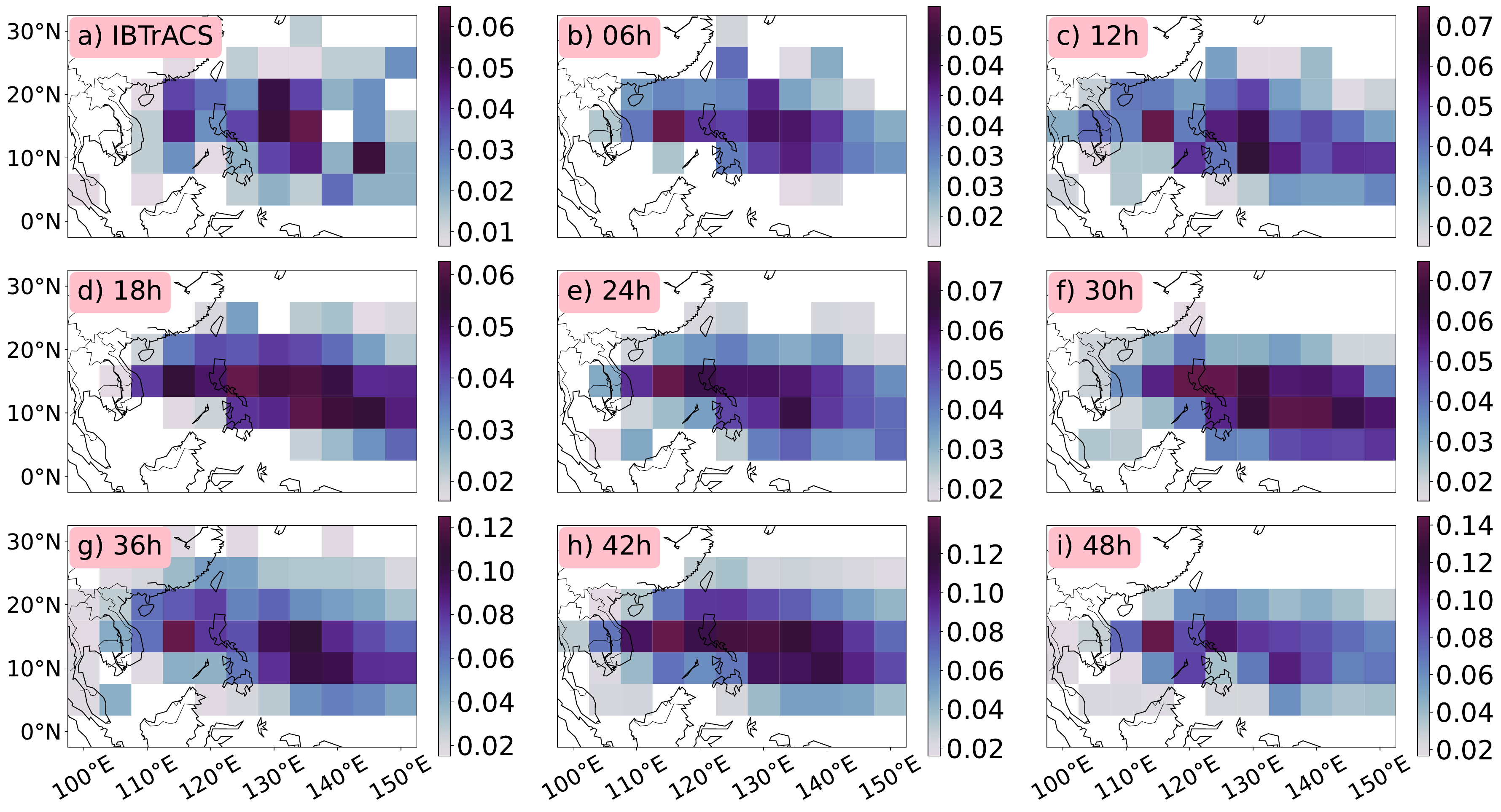}
\caption{a) The spatial distribution of the observed TCG density (shaded) during the 2017-2022 period as obtained from the best track; (b)-(i) 5-year average of TCG probability prediction that is obtained from the ResNet-18 model with different data enrichment windows from 6-48 hours during the same test period as in (a). Note that different shading scales are used for different data enrichment windows so that one can better see the contrast between the areas of maximum probability for TCG predicted by the ResNet-18 model.}
\label{fig:spatialTCGdist}
\end{center}
\end{figure}

To further look into the spatial distribution of TCG climatology by the ResNet-18 model, Figure \ref{fig:spatialTCGdist} shows the horizontal map of the predicted TCG probability over the entire WNP basin, using the DD strategy with data enrichment windows from 6 to 48 hours. Here, the shading in each $5^\circ \times 5^\circ$ box represents the averaged probability of positive TCG predictions from the ResNet-18 model over the test period from 2017-2022. For the observed TCG, the shading denotes the actual number of TCG events occurs in each box, divided by the total number of TCGs during the 2017-2022 period (also commonly known as TCG density in literature). Thus, the predicted and observed TCG distribution are proportional to each other, and can serve as a metrics to evaluate our DL model's performance from a different angle.  %although they may differ by a multiplicative factor that is proportional to the total number of TC seeds each year.     

Overall, the ResNet-18 model could again reproduce well the TCG distribution in the WNP basin during the 2017-2022 period, with clear TCG hotspots in both the East Philippine Sea and the South China Sea (SCS). For long enrichment windows between 18-42 hours, note that the ResNet-18 model tends to extend the TCG region too far east of the East Philippine Sea as compared to the observation. However, the overall spatial distribution of TCG probability is still distinct and concentrated in the central SCS, the eastern Philippine Sea, and the Vietnam coastal region. For the shorter data enrichment window of 6-12 hours, the model could also provide a reasonable fit as compared to the observed distribution (Fig. \ref{fig:spatialTCGdist}b), despite not being as good as the longer windows or the seasonal distribution in Fig. \ref{fig:seasonalTCGdist}.  

It is worth noting that while the predicted TCG probabilities consistently peak in the eastern Philippine Sea, some localized maxima in the central SCS and along the Vietnamese coastline lack consistency across data enrichment windows. This variability reflects the inherent challenge of detecting early TCG signals in the SCS, which are typically weak and highly variable. As a result, the ResNet-18 model, when optimized based on the F1 score over the broader WNP basin, struggles to capture these localized signals effectively. Unfortunately, the amount of available TCG data in the SCS region alone is insufficient for meaningful DL model training, making it difficult to resolve such inconsistencies given the current limitations of climate datasets. Because of this, we could only reconstruct the TCG climatology using the DD strategy as presented in this subsection, even though the PD strategy is more directly applicable for real-time forecasting purposes.

Aside from these local issues, the ability of our DL model in capturing the broad spatiotemporal patterns of TCG suggests that atmospheric signals associated with TCG become increasingly detectable when more relevant information is incorporated. From a practical standpoint, these results are very significant, because not only do they validate the performance of our DL model, but they also demonstrate that TCG can be predicted from large-scale environmental information, even at a spatial resolution of 0.5 degrees. This result has two important consequences: 1) as long as climate models can reliably simulate the large-scale environment, it is possible to learn TCG patterns with DL models and derive TCG climatology without resorting to the more computationally expensive high-resolution dynamical downscaling, and 2) changes in TCG climatology can be captured through changes in large-scale environments, which are generally much more robust and reliable in climate projections than individual storm-scale features.

%%%%%%%%%%%%%%%%%%%%%%%%%%%%%%%%%%%%%%
\subsection{Sensitivity analyses}
\label{subsec:Sensitivity-Features}
%%%%%%%%%%%%%%%%%%%%%%%%%%%%%%%%%%%%%%
One critical question in developing a DL model for applications when the training data does not contain sufficient information is how to make full use of the data to optimize the model's performance. This process, known as feature engineering, becomes more significant when one tries to understand why we get what we see from DL model's output. Instead of running a DL model as a black box with all possible input data channels, exploring the importance of different input data can help better understand the role of different physical information in the model prediction that we wish to examine from a physical standpoint.

This subsection provides several additional analyses that employ a different set of input channels to see how effectively the ResNet-18 model could perform with limited information. Specifically, we examine two additional analyses that 1) use a set of features known to be of importance for TCG from previous studies, and 2) apply an automatic feature filter based on the rank of input channels. While this is often considered to be a part of model tuning, we treat them separately in this study. The choosing the right input channels will have significant implications in our further understanding of TCG processes.

\begin{table}[t!]
    \centering
    \begin{tabular}{|c|c|c|c|c|c|}
    \hline
        \multirow{2}{*}{ID} & \multirow{2}{*}{Name of Features} & \multicolumn{2}{c|}{Past Domain} &  \multicolumn{2}{c|}{Dynamic Domain} \\ 
         & & Engineering & Ranking & Engineering & Ranking \\ \hline
         1 & QL &  & \makecell[c]{400, 700, 825,\\ 900,\\ 950}&  &\makecell[c]{100, 1000. 150,\\ 200, 300, 400,\\ 500, 600, 700,\\ 800, 875, 900,\\ 950, 975}  \\
         2 & H  & 500 & 200, 925 & 500  & 100, 550, 950 \\
         3 & QI &  & \makecell[c]{250, 450, 600,\\ 800, 900, 925,\\ 950, 1000}  &  & \makecell[c]{100, 1000, 150,\\ 500, 600, 700,\\ 900} \\
         4 & OMEGA & 500 & 450, 875 & 500 & \makecell[c]{100, 150, 250,\\ 600, 925, 1000} \\
         5 & T & 500, 900 & 725 & 500, 900 &150, 200, 900  \\
         6 & U & 200, 800 & 825, 1000 & 200, 800 & 1000, 200, 550 \\
         7 & V & 200, 800 & 150, 550  & 200, 800 & \makecell[c]{100, 1000, 150,\\ 400, 600} \\
         8 &  RH & 750 &950  &750  & \makecell[c]{100, 200, 400,\\ 700, 825, 875,\\ 925, 1000} \\
         9 & QV &  &  &  & 100, 150, 900 \\
         10 & VOR & 200, 700, 900 &  & 200, 700, 900 &  \\
         11 & DIV & 200  &  & 200 &  \\ \hline
    \end{tabular}
    \caption{List of features selected using the feature engineering and feature ranking filter approach as obtained for each labeling strategy during the training period.}
    \label{tab:featureSelect}
\end{table}

For the first approach (hereafter referred to as feature engineering), input channels are selected based on their importance as obtained from previous observational and modeling studies \citep[See, e.g.,][]{Gray1968, RiehlMalkus1958, Yanai1964, ZhangBao1996a, BisterEmanuel1997, RitchieHolland1997, Simpson_etal1997, Molinari_etal2000, Nolan_etal2007, NguyenKieu2024, Kieu_etal2023}. Since not all meteorological data have strong control on TCG, only a subset of atmospheric variables that display the strongest TCG signals will be taken into consideration. Note that data on adjacent pressure levels are often correlated with each other. Therefore, our feature engineering approach will be applied to a few layers that can be categorized into low, middle and high pressure levels as shown in Table \ref{tab:featureSelect} instead of every single levels. %In addition to existing features, vorticity and divergence are also useful for predicting the formation, derived from the U and V wind components.

For the feature filtering approach (hereafter feature ranking), data channels are automatically selected by their contribution to the prediction score instead of depending on specialized knowledge, as for the feature engineering approach. This can be done by ranking all input features in terms of the mean activation value of the first filter from our best-tuned ResNet-18 model. Selection is then proceeded iteratively as follows. First, the channel with the highest score in the scoreboard is chosen and removed from the pool. Next, we eliminate any remaining layers that are highly correlated with the chosen channel (i.e., exceeding a Pearson $R$ threshold). Finally, the progress is continued until all remaining features are either selected from their score or discarded, depending on the threshold that is used to stop the selection.

Table~\ref{tab:featureSelect} compares the features after applying the two feature selection methods on the MERRA-2 data. If is of interest to note that the selected features between these two methods share a great number of overlaps, indicating that previous findings on TCG factors such as vertical wind shear (zonal wind components at 800 and 200-hPa level), low-level moisture (900-700 hPa relative humidity, RH), mid-level vertical motion (OMEGA at 500 hPa), or low-to-mid level temperature all play an important role in TCG prediction. 

In addition, we observe that feature ranking tends to select a broader range of pressure levels than feature engineering. For example, NK2024's feature engineering uses RH at 750 hPa, while feature ranking captures a larger number of levels for the DD strategy (see last column in Table \ref{tab:featureSelect}). Likewise, zonal wind components in feature engineering require only 200 and 800 hPa, but the feature ranking shows several levels at 1000, 600, 400, 150, and 100 hPa. However, the fact that both feature engineering and feature ranking could share a large group of features apparently indicate that the DL model is capable of learning large-scale environments correctly for the TCG problem. From this regard, our DL model not only justifies the use of the feature engineering method in previous studies, but also presents a way to help enhance our understanding of the key environments governing TCG when the ranking is expanded to include more environmental factors.   

Among all selected features, we should note that there are several features from the feature ranking method that are not included in the feature engineering such as the specific liquid (QL) and ice content (QI), which represent the cloud information. In their early study, NK2024 did not include these features as they are linked to vertical motion and relative humidity via CAPE channel. Similarly, the specific humidity (QV) is just an equivalent representation of RH and so it is not included in feature engineering. These variables, however, emerge in the feature ranking, probably because of their stronger roles in capturing TCG during the training, even though they do not provide any new physical implications. Note also that the original MERRA-2 data does not contain some fields such as vorticity or divergence used in NK2024. Thus, several features in NK2024's study cannot be obtained from the ranking feature method shown in Table \ref{tab:featureSelect}.

%Comparing the PD and DD approaches, we observe further that the DD approach includes more upper-level levels (e.g., 100 hPa), while the PD approach tends to focus more on mid-to-lower atmospheric layers. This may reflect the nature of each approach, where the PD emphasizes the accumulated conditions and the DD requires capturing rapid temporal changes and atmospheric dynamics relevant to TCG. 

\begin{figure}[t!]
\begin{center}
\includegraphics[width=0.9\linewidth]{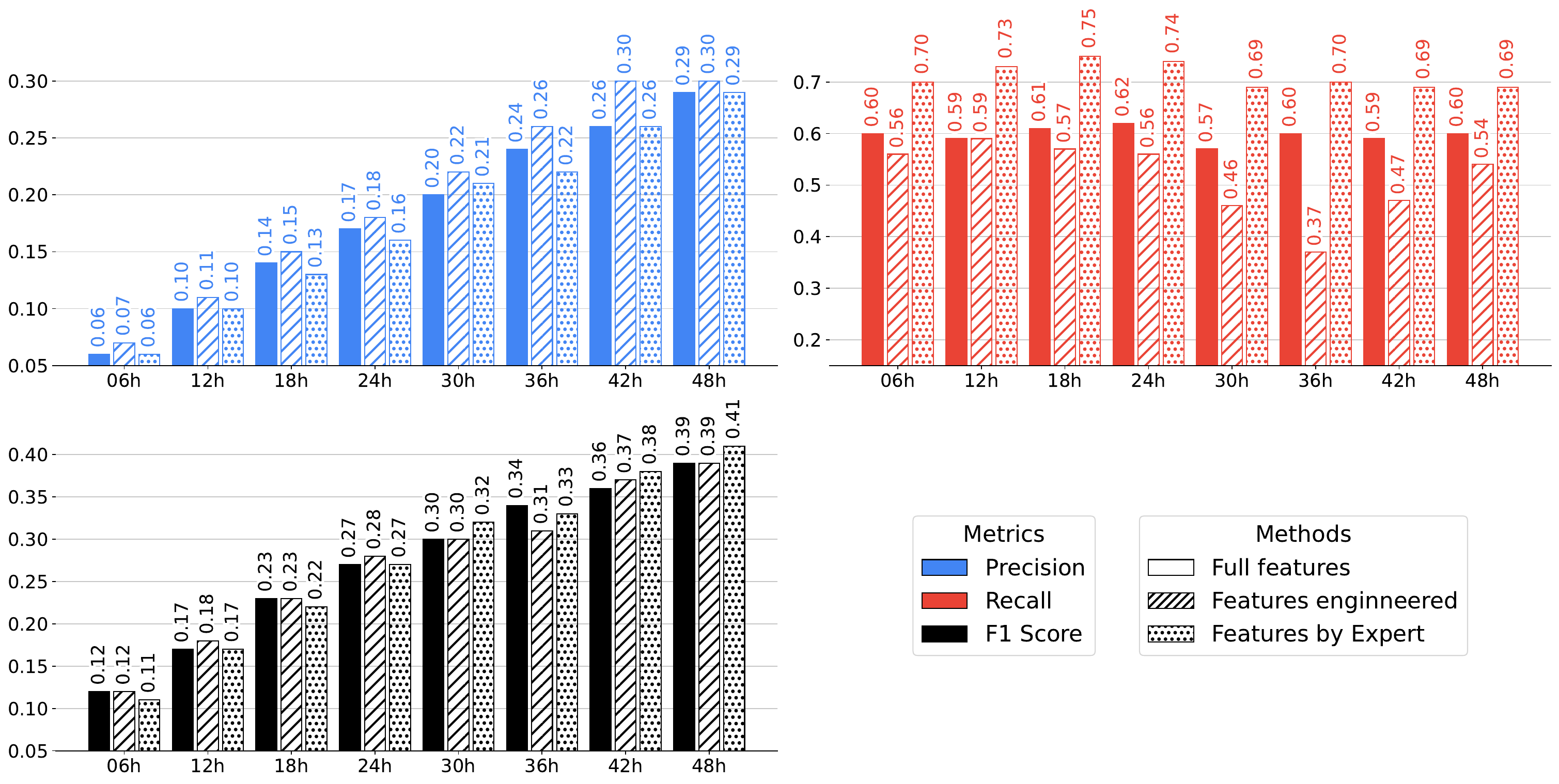}
\caption{Evaluation of the model performance for several different feature selection methods including all features (solid color columns), 13 selected features based on feature engineering in \cite{NguyenKieu2024} (striped columns), and feature ranking of top 10\% (dotted columns) using the Past Domain strategy for a) Precision, b) Recall, and c) F1 score.}
\label{fig:feaSelectPD}
\end{center}
\end{figure}

\begin{figure}[t!]
\begin{center}
\includegraphics[width=0.9\linewidth]{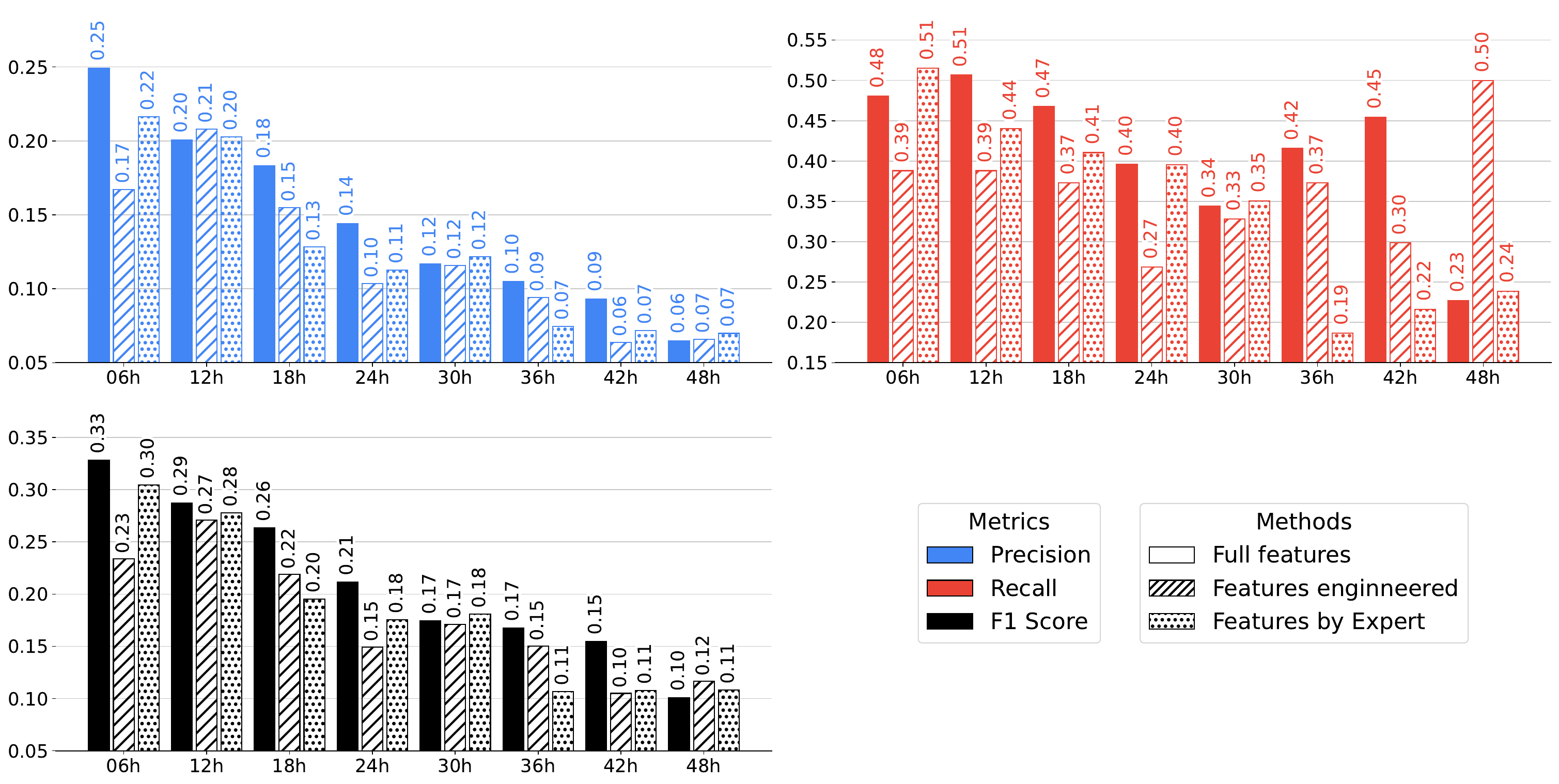}
\caption{Similar to Fig. \ref{fig:feaSelectPD} but for the Dynamic Domain strategy.}
\label{fig:feaSelectDD}
\end{center}
\end{figure}

In terms of the model performance, Figures \ref{fig:feaSelectPD}-\ref{fig:feaSelectDD} compare the F1, $P$, and $R$ scores among the full-feature, feature engineering, and feature ranking methods for both PD and DD labeling strategies. Among these feature selection methods, the feature ranking appears to deliver the most stable and effective results for the PD labeling strategy, particularly for long enrichment windows. The feature engineering and full-feature methods also perform well, but with slightly lower overall performance as compared to the feature engineering. 

In contrast, the DD labeling strategy shows a declining trend in performance, with $P, R$ and F1 score all decreased over longer time windows when using the all-feature method. However, the feature engineering method still maintains relatively higher and more stable $R$ values, indicating its potential utility in capturing more relevant signals in predicting TCG spatial distribution without the need of all data channels. 

\begin{figure}[t]
\begin{center}
\includegraphics[width=0.8\linewidth]{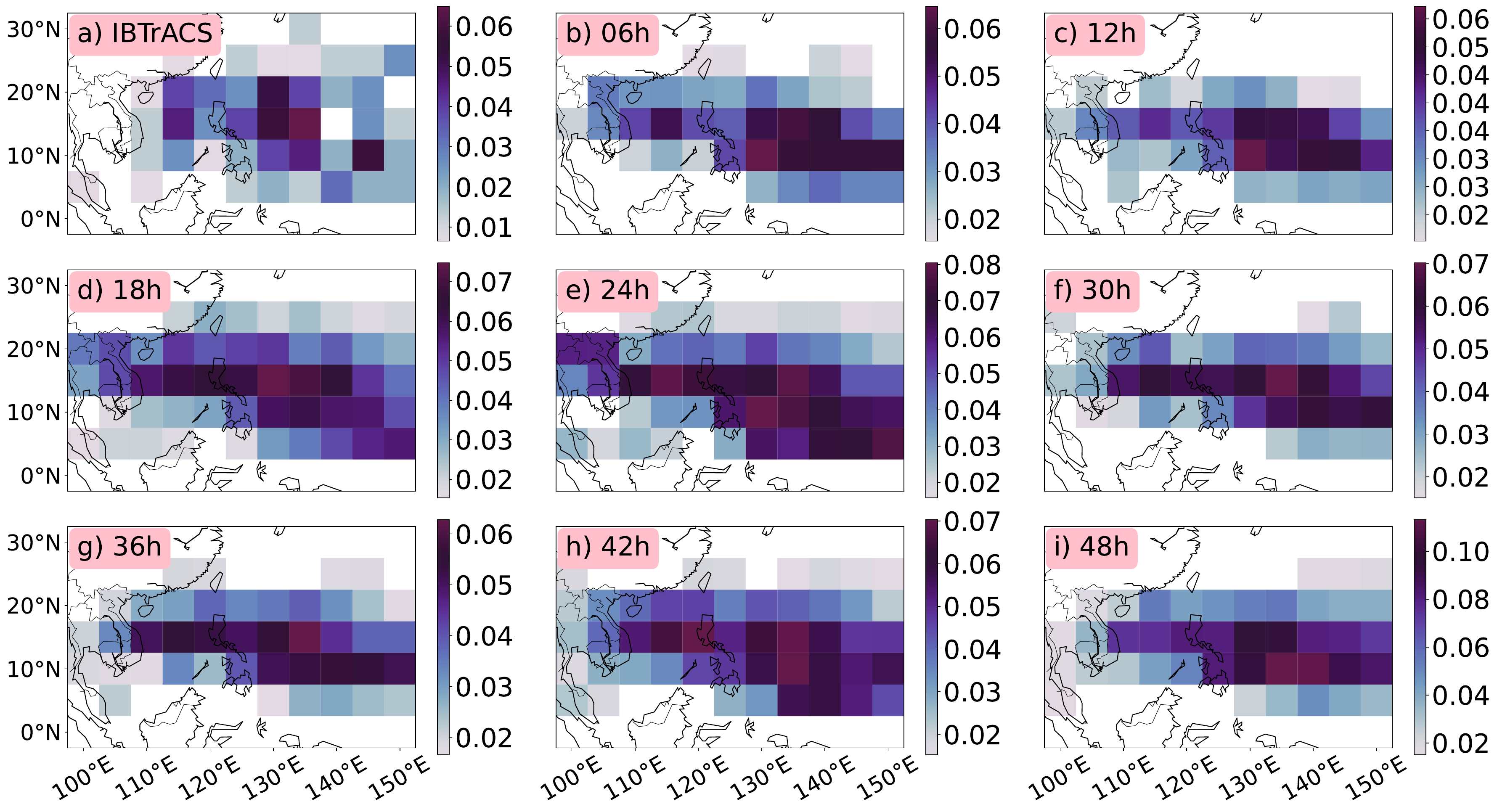}
\caption{The spatial distributions of TCG climatology on map using the feature engineering approach.}
\label{fig:feature_engineering}
\end{center}
\end{figure}

\begin{figure}[t]
\begin{center}
\includegraphics[width=0.8\linewidth]{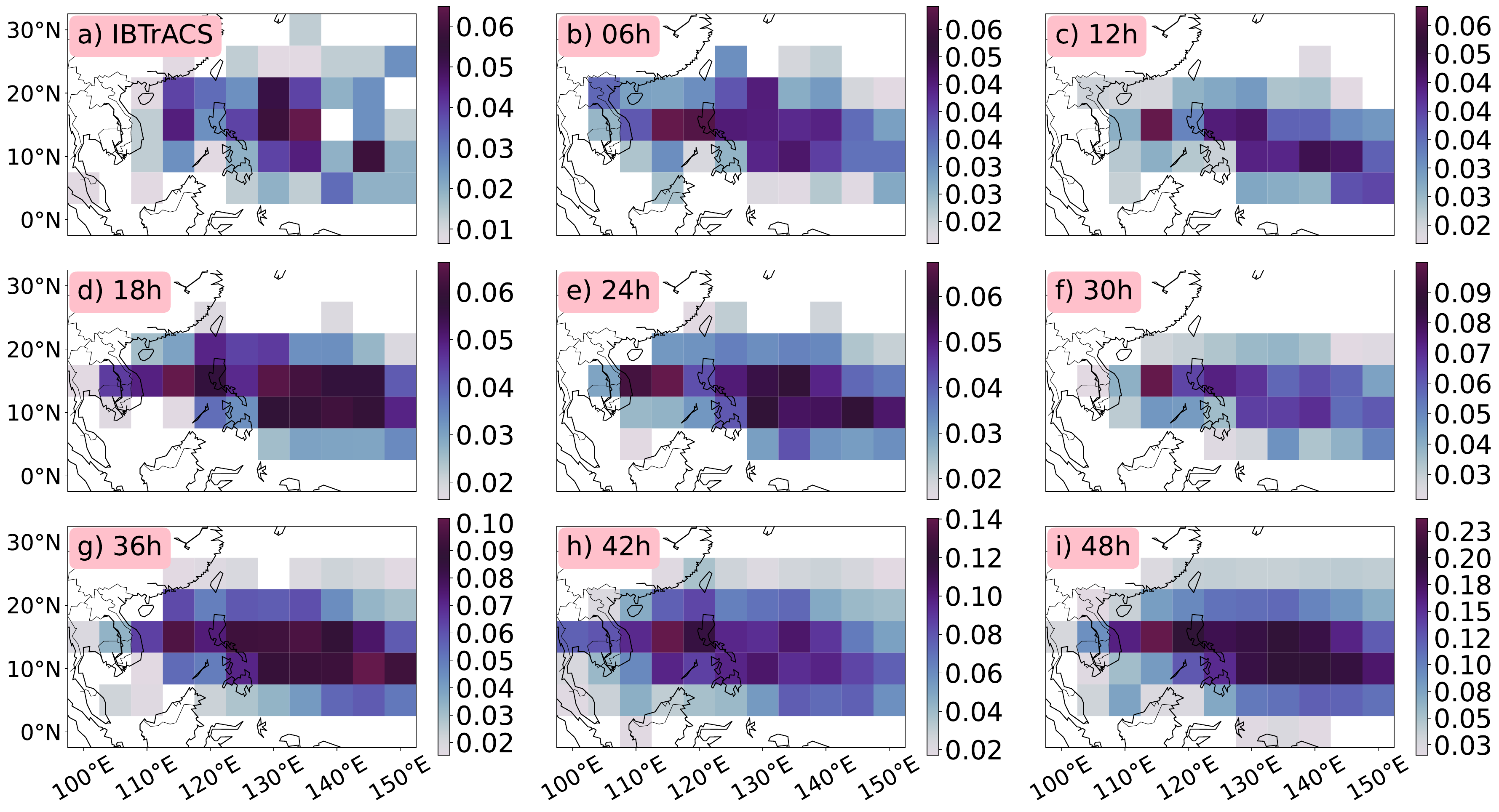}
\caption{The spatial distributions of TCG climatology on map using the automatic feature ranking approach.}
\label{fig:feature_ranking}
\end{center}
\end{figure}

Regardless of the labeling strategy, one can see that both feature engineering and feature ranking provide similar performance for a range of data enrichment windows, despite much fewer features than the full-feature method. For a long window of 36-48 hours, feature engineering and feature ranking could deliver even better performance for the PD approach, thus confirming that predicting TCG would mostly rely on a subset of variables/channels instead of full data at all levels. In fact, examining the spatial distributions of TCG climatology from both feature engineering and feature ranking (Figs \ref{fig:feature_engineering}-\ref{fig:feature_ranking}) shows little change in the overall patterns as compared to the full-feature output (cf. Fig. \ref{fig:spatialTCGdist}). The consistency among all feature selection methods is also captured for the seasonal TCG density (not shown), thus giving us some foundation for further understanding and predicting TCG based on a set of selected features for future DL improvement or implementations.       

\section{Conclusion}
\label{sec:Conclusion}
%%%%%%%%%%%%%%%%%%%%%%%%%%%%%%%%%%%%%%
In this study, we presented a deep learning (DL) framework for reconstructing the climatology of tropical cyclone genesis (TCG) from climate reanalysis datasets. Recognizing that the definition of TCG climatology may vary depending on specific purposes and practical needs, our DL approach was designed and evaluated from multiple perspectives, based on different TCG labeling strategies for model training. Due to the limited number of TCG events available for training, we also implemented different data enrichment and feature selection methods to optimize our DL model for both TCG climatology reconstruction and potential prediction tasks.

Using the MERRA-2 reanalysis as the training dataset and the ResNet-18 architecture as our DL model, we demonstrated that ResNet-18 exhibits promising capability in detecting TCG from climate data at 0.5$^\circ$ resolution. Although the F1 score for TCG prediction remains relatively low, which is partly due to the inherent low predictability of TCG, the limited TCG samples, and related information available in MERRA-2, we showed that the F1 score can be improved through appropriate hyperparameter tuning, labeling strategies, class weighting, and feature selection. 

Comparing the reconstructed TCG seasonality and spatial distribution to the best track during the 2017-2022 period showed several noteworthy results. First, our ResNet-18 design could reproduce the seasonality of monthly TCG frequency, with double peaks of TCG frequency in August and October as well as the inactive period from January to May, consistent with those obtained from observations. Second, ResNet-18 could recover also the spatial distribution of TCG climatology, with main areas in the Eastern Philippine Sea and SCS. While there are some fluctuations in the TCG distribution for the areas along the coastal regions of the WNP basin, the overall well-recovered map of TCG in the WNP basin indicates that large-scale environments from the MERRA-2 dataset contain some important hidden signals of TCG processes that DL models can be trained and learn.  

Further sensitivity experiments with different feature selection methods revealed that reconstructing TCG from any climate datasets is possible due to the existence of some key channels that contain the required TCG signals for DL models to learn. Specifically, our use of feature engineering based on a set of features reported in previous studies and feature ranking that filters input features based on the model impacts both capture some common data channels (variables) needed for TCG processes. Several of these key features obtained from the MERRA-2 dataset include vertical wind shear, low-to-mid level moisture, mid-level vertical motion, or mid-level geopotential height. The feature ranking method could detect some additional features such as high-level liquid or ice content that may include some cloud signal information beyond what used in previous models.     

The results from these feature-sensitivity experiments support that combining expert knowledge with automated feature engineering can help enhance feature representation when training data is not sufficient or when running full features is too costly. Moreover, while feature engineering helps avoid overfitting and promotes interpretability, the feature ranking approach can help uncover hidden signals beyond what known from previous studies. For our TCG problem here, at least both the feature engineering and feature ranking approaches could share similar factors, which help build more confidence in using those input features to understand the variability of TCG climatology. Future improvements could involve integrating approaches using ranking mechanisms, attention-based models, or dimensionality reduction to retain a compact yet informative feature set for TCG prediction tasks. 

%conducted an extensive investigation into how various data resampling strategies and class weighting schemes influence model performance across different forecast horizons in an imbalanced classification setting. Through a series of controlled experiments, we evaluated the effects of multiple undersampling ratios combined with both dynamic and balanced weighting techniques, using standard classification metrics such as precision, recall, and F1 score.

Along with the primary focus on reconstructing TCG climatology, this study suggests that our approach holds potential for real-time TCG prediction in operational settings. Depending on the specific forecasting objective, i.e., whether predicting TCG at a fixed location (as in the PD labeling strategy) or forecasting the spatial distribution of TCG at a given time (as in the DD labeling strategy), a DL model can be designed and trained for each task separately. While the relatively low F1 scores observed for both strategies indicate that real-time forecasts of TCG from climate or global model output would be subject to considerable uncertainty, our DL approach offers a valuable, independent alternative to traditional physical-based models, with the capability to provide early warnings of TCG 1–3 days in advance. In this context, our work contributes to a deeper understanding of TCG processes and offers another practical guidance for improving DL model in real-time applications, particularly for extreme events where minority classes are the main focus.

Despite promising capabilities, this study reveals also several key challenges in applying DL models to TCG research. First, DL performance is highly sensitive to data preprocessing methods, particularly in labeling positive and negative TCG events, an issue that is exacerbated by the limited number of observed TCG occurrences. Second, the pronounced class imbalance between positive and negative TCG labels during training remains a significant barrier. Addressing this challenge requires a careful integration of undersampling techniques, data augmentation strategies, or dynamic class weighting to ensure more robust and consistent performance across evaluation metrics. These challenges highlight the complexity of DL applications to extreme weather events. Moving forward, we aim to refine these approaches to extend TCG climatology reconstructions further back into the pre-satellite era for which we will present in our upcoming study. 

% %%%%%%%%%%%%%%%%%%%%%%%%%%%%%%%%%%%%%%
% \section{Acknowledgements}
% \label{sec:Acknowledgements}
% %%%%%%%%%%%%%%%%%%%%%%%%%%%%%%%%%%%%%%

% This research was funded by Vingroup Innovation Foundation (VINIF, code VINIF.2023.DA019).

% \noindent{\em Funding info:}
% VINIF, Grant Number VINIF.2023.DA019; NSF, Grant/Award Number: 135999

% %%%%%%%%%%%%%%%%%%%%%%%%%%%%%%%%%%%%%%
% \section{Conflict of Interest}
% %%%%%%%%%%%%%%%%%%%%%%%%%%%%%%%%%%%%%%
% The authors declare no conflict of interest.

%% The following commands are for the statements about the availability of data sets and/or software code corresponding to the manuscript.
%% It is strongly recommended to make use of these sections in case data sets and/or software code have been part of your research the article is based on.

\codeavailability{The complete source code and pipeline of the \mName{} framework used for reconstructing TCG climatology in this study are publicly available in \cite{le2025zenodo}, \url{https://doi.org/10.5281/zenodo.16741501}. The repository includes all scripts for data preprocessing, model training, evaluation metrics (including sensitivity analysis), and visualization. Our code structure is designed in a way to support not only the reproducibility of results herein but also further experimentation with different climate datasets or TCG-related tasks. A detailed user guide for running the full workflow can be found in our repository.} %% use this section when having only software code available

\dataavailability{This research explores two publicly available datasets including: the NASA’s Modern-Era Retrospective analysis for Research and Applications Version 2 (MERRA-2, URL: \url{https://disc.gsfc.nasa.gov/datasets?project=MERRA-2})  and the International Best Track Archive for Climate Stewardship (IBTrACS, URL: \url{https://www.ncei.noaa.gov/products/international-best-track-archive}).} %% use this section when having only data sets available
% The former is a global atmospheric reanalysis dataset that combines satellite observations with a weather model to provide consistent records of the atmosphere, land, and oceans from 1980 to the present whilst the latter is a global database that brings together tropical cyclone track and intensity information from multiple meteorological agencies into a single, consistent archive.

% \codedataavailability{TEXT} %% use this section when having data sets and software code available

% \sampleavailability{TEXT} %% use this section when having geoscientific samples available

% \videosupplement{TEXT} %% use this section when having video supplements available

% \appendix
% \section{}    %% Appendix A

% \subsection{}     %% Appendix A1, A2, etc.

\noappendix       %% use this to mark the end of the appendix section. Otherwise the figures might be numbered incorrectly (e.g. 10 instead of 1).

%% Regarding figures and tables in appendices, the following two options are possible depending on your general handling of figures and tables in the manuscript environment:

%% Option 1: If you sorted all figures and tables into the sections of the text, please also sort the appendix figures and appendix tables into the respective appendix sections.
%% They will be correctly named automatically.

%% Option 2: If you put all figures after the reference list, please insert appendix tables and figures after the normal tables and figures.
%% To rename them correctly to A1, A2, etc., please add the following commands in front of them:

\appendixfigures  %% needs to be added in front of appendix figures

\appendixtables   %% needs to be added in front of appendix tables

%% Please add \clearpage between each table and/or figure. Further guidelines on figures and tables can be found below.

\authorcontribution{
\textbf{Duc-Trong Le}: Formal analysis, Conceptualization, Methodology, Validation, Supervision, Writing-review\&editting. \textbf{Tran-Binh Dang}: Investigation, Software, Validation, Writing-original draft. \textbf{Anh-Duc Hoang Gia}: Methodology, Software, Writing-original draft. \textbf{Duc-Hai Nguyen}: Software, Validation. \textbf{Minh-Hoa Tien}: Software, Validation. \textbf{Xuan-Truong Ngo}: Software, Validation. \textbf{Quang-Trung Luu}: Software, Validation, Writing-review\&editting. \textbf{Quang-Lap Luu}: Software, Validation. \textbf{Tai-Hung Nguyen}: Validation, Supervision. \textbf{Thanh T.N. Nguyen}: Formal analysis, Conceptualization, Funding acquisition,
Methodology, Supervision. \textbf{Chanh Kieu}: Conceptualization, Methodology, Validation, Supervision, Writing-review\&editting.
} %% this section is mandatory

\competinginterests{The contact author has declared that none of the authors has any competing interests.} %% this section is mandatory even if you declare that no competing interests are present

% \disclaimer{TEXT} %% optional section

\begin{acknowledgements}
 This research is supported by Vingroup Innovation Foundation (VINIF, code VINIF.2023.DA019) and the U.S. National Science Foundation (NSF,  Grant/Award Number 135999)
\end{acknowledgements}

%% REFERENCES

%% The reference list is compiled as follows:

% \begin{thebibliography}{}

% \bibitem[AUTHOR(YEAR)]{LABEL1}
% REFERENCE 1

% \bibitem[AUTHOR(YEAR)]{LABEL2}
% REFERENCE 2

% \end{thebibliography}

\bibliographystyle{copernicus}
\bibliography{reference1,reference2,reference3}

\end{document}